\documentclass[a4paper,11pt]{article}
\pdfoutput=1 

\usepackage{jheppub} 

\usepackage[T1]{fontenc} 

\title{\boldmath $T\bar{T}$-like Flows in Non-linear Electrodynamic Theories and S-duality}

\author[a]{H. Babaei-Aghbolagh,}
\author[b,c]{Komeil Babaei Velni,}
\author[d]{Davood Mahdavian Yekta,}
\author[a]{and H. Mohammadzadeh}

\affiliation[a]{Department of Physics, University of Mohaghegh Ardabili,
P.O. Box 179, Ardabil, Iran}
\affiliation[b]{Department of Physics, University of Guilan, P.O. Box 41335-1914, Rasht, Iran}
\affiliation[c]{School of Physics and School of Particles and Accelerators,
Institute for Research in Fundamental Sciences (IPM), P.O. Box 19395-5531, Tehran, Iran}
\affiliation[d]{Department of Physics, Hakim Sabzevari University, P.O. Box 397, Sabzevar, Iran}
\emailAdd{h.babaei@uma.ac.ir}
\emailAdd{babaeivelni@guilan.ac.ir}
\emailAdd{d.mahdavian@hsu.ac.ir}
\emailAdd{mohammadzadeh@uma.ac.ir}

\abstract{We investigate the $T\bar{T}$-like flows for non-linear electrodynamic theories in $D(=\!\!2n)$-dimensional spacetime. Our analysis is restricted to the deformation problem of the classical free action by employing the proposed $T\bar{T}$ operator from a simple integration technique. We show that this flow equation is compatible with $T\bar{T}$ deformation of a scalar field theory in $D\!=\!2$ and of a non-linear Born-Infeld type theory in $D\!=\!4$ dimensions. However, our computation discloses that this kind of $T\bar{T}$ flow in higher dimensions is essentially different from deformation that has been derived from the AdS/CFT interpretations. Indeed, the gravity that may be exist as a holographic dual theory of this kind of effective Born-Infeld action is not necessarily an AdS space. As an illustrative investigation in $D\!=\!4$, we shall also show that our construction for the $T\bar{T}$ operator preserves the original $SL(2,R)$ symmetry of a non-supersymmetric Born-Infeld theory, as well as $\mathcal{N}=2$ supersymmetric model. It is shown that the corresponding $SL(2,R)$ invariant action fixes the relationship between the $T\bar{T}$ operator and quadratic form of the energy-momentum tensor in $D\!=\!4$.}

\begin{document}
\maketitle
\flushbottom

\section{Introduction}
\label{sec1}
One of the fundamental concepts in any quantum field theory (QFT) is the implication of the renormalization group approach that helps us to better understand the space of QFTs. About two decades ago, it has been proposed by Zamolodchikov \cite{Zamolodchikov:2004ce} that one can construct this space by starting from a fixed point of the renormalization group and perturbing the free or exactly solvable action of a 2D QFT by an integrated local operator. In this regard, using a relevant operator, the renormalization group flows to some IR fixed points while for an irrelevant operator, the UV properties of the theory will change.

In ref.~\cite{Smirnov:2016lqw}, Smirnov and Zamolodchikov discovered a general class of exactly solvable irrelevant deformations of 2D conformal field theories (CFT) known as the $T\bar{T}$ deformation (see also \cite{Cavaglia:2016oda}). Even though the $T\bar{T}$ deformations provide laboratories for investigating various aspects of field theories \cite{Guica:2017lia,Chen:2018keo,Aharony:2018bad,Cardy:2018sdv,Datta:2018thy,Bonelli:2018kik,Brennan:2019azg}, but most of the attempts have an interpretation from the AdS/CFT correspondence point of view. It was conjectured in refs.~\cite{McGough:2016lol,Kraus:2018xrn} that the $T\bar{T}$ deformed CFT can be interpreted as the holographic dual of a finite patch (a finite radial cutoff) of asymptotically AdS spacetime. This result is equivalent to the statement that the holographic dual at finite radius should be interpreted as a deformed CFT \cite{Heemskerk:2009pn}. Other related interesting developments can be found in refs.~\cite{Giveon:2017myj,Giribet:2017imm,Aharony:2018vux,Donnelly:2018bef,Hartman:2018tkw,Conti:2018tca}.

It has been shown in ref.~\cite{McGough:2016lol} that many interesting physical observables can be computed explicitly from such non-local UV theory. Taking into account a $T\bar{T}$ operator (an irrelevant operator given by the product of $T=T_{zz}$ and $\bar{T}={T}_{\bar{z}\bar{z}}$ components of the energy-momentum (EM) tensor), we have a Lorentz invariant deformation as $S_{QFT}=S_{CFT}+S_{\lambda}$, where $S_{\lambda}=\lambda\int d^2 x \, T\bar{T}$. In this respect, some theories emerge by ${d S_{QFT}}/{d\lambda}=\int d^2 x \, (T\bar{T})_{\lambda}$ for finite $\lambda$.

In 2D, the $T\bar{T}$ operator is well-defined quantum mechanically by examining the energy spectrum (it is free of short distance singularities) \cite{McGough:2016lol}. In definition of the $T\bar{T}$ deformation proposed by Zamolodchikov that the variation of the Lagrangian with respect to the deformation parameter equals the determinant of the deformed EM tensor, all objects are quantum renormalized and UV finite. Nevertheless, one can treat the resulting Lagrangian as a classical one, giving rise to classical fields which is in accordance with some features of the corresponding models.

The generalization to higher dimensions from the field theory point of view has also been proposed in refs.~\cite{Cardy:2018sdv,Bonelli:2018kik}. However, in general dimensions the operator does not share all the special properties uncovered by Zamolodchikov in two dimensions. In fact, the $T\bar{T}$ deformation at higher than two dimensions is not unique. It was proposed in \cite{Cardy:2018sdv} to use $(det\, T)^{1/\alpha}$ with $\alpha\!=\!D-2$ and in \cite{Bonelli:2018kik} it was further generalized to different values of $\alpha$, or even additional deformations are required if one includes bulk scalars and gauge fields.

Considering the approach in \cite{Kraus:2018xrn}, a generalized $T\bar{T}$ deformation in D dimensions was proposed in ref.~\cite{Taylor:2018xcy} in the context of holographic realization of $AdS_{D+1}$ spacetime. As the main purpose of this paper, we recast the $T\bar{T}$ operator given in ref.~\cite{Smirnov:2016lqw} for general dimension $D$ as follows
\begin{equation}
\label{TT1}
{O}_{T^2}^{[r]}=T_{\mu\nu}T^{\mu\nu}-r {T_{\mu}}^{\mu} {T_{\nu}}^{\nu},
\end{equation}
where $r=1/(D-1)$ from holographic calculations \cite{Taylor:2018xcy}.

According to the AdS/CFT conjecture \cite{Maldacena:1997re}, the type IIB superstring theory on $AdS_5\times S^5$ is daul to $\mathcal{N}\!=\!4$ supersymmetric Yang-Mills theory in four dimensions. Indeed, the latter is a non-abelian $SU(N)$ guage theory with large $N$ which lives on $N$ coincident $D3$-branes in string theory. However, the effective action of a single brane is described by an abelian theory which specially in $D\!=\!4$ is a Born-Infeld (BI) theory and its gravity dual is not necessarily an $AdS$ theory. In this respect, the $T\bar{T}$ operator in \eqref{TT1} which has been found by using the Lax pair operator in ref.~\cite{Conti:2018jho} for BI theory of Maxwell field in $D\!=\!4$, resembles a higher dimensional generalization of $T\bar{T}$ operator except for the factor $r = 1/2$ instead of $r = 1/(D-1) = 1/3$. This incompatibility motivates us strongly to investigate the $T\bar{T}$-like flows in higher dimensional gauge theories. 

Though the gravity duals of field theories which are not necessarily CFT are not clear, but recently there are attempts to generalize the $T\bar{T}$ operator for these QFTs in $D\!=\!4$. For example, the extension of this operator for $\mathcal{N}\!=\!1$ supersymmetric BI theory in \cite{Ferko:2019oyv}, the emergent gravity from hidden QFT coupled to the standard model via $T\bar{T}$ deformation in \cite{Betzios:2020sro}, the $T\bar{T}$-like deformation of the Skyrme action as generalizations of the Heisenberg model for nucleon-nucleon scattering in \cite{Nastase:2021mgy} and the $T\bar{T}$ deformation of non-Lorentz invariant models, such as the nonlinear Schr\"{o}dinger equation, the Landau-Lifshitz equation and the Gardner equation in \cite{Frolov:2021oav}.

In particular, we study a class of QFTs which are described by non-linear electrodynamic BI type Lagrangian in $D\!=\!2n\!=\!2(p+1)$ dimensions (for even $n$) and constructed from $p$--form field $A_{\mu_1 \dots \mu_p}$ with strength $F_{\mu_1 \dots \mu_n}=n\, \partial_{[\mu_1}A_{\mu_2 \dots \mu_n]}$.
The dynamics of this field is described by a Lorentz invariant local Lagrangian $L(F)$ which entails the field equations
\begin{equation}
\label{afe} \partial_{\mu_1}\frac{\partial L(F)}{\partial F_{\mu_1 \dots \mu_n}}=0.
\end{equation}
The proposed $T\bar{T}$ operator deserves the results obtained in refs.~\cite{Smirnov:2016lqw,Conti:2018jho} for 2D QFTs and generalized 4D BI models \cite{Conti:2018jho}, respectively.

In general, the EM tensor of any QFT can be obtained as the variational derivative of the Lagrangian with respect to the gravitational field (i.e. the spacetime metric). However, it has been found in refs.~\cite{Gibbons:1995ap,Gaillard:1997rt} that for a given Lagrangian in $D\!=\!2n$ dimensions which depends only on the self-interacting, completely antisymmetric field strength, such as non-linear electrodynamic theories, the EM tensor can be written as
\begin{equation}
\label{EMT1}
T_{\mu\nu}=g_{\mu\nu} L(F)+\frac{1}{(n-1)!}(G_{\mu}F_{\nu}),
\end{equation}
where $(G_{\mu}F_{\nu})=G_{\mu \alpha_1 \dots \alpha_{n-1}}\,{F_{\nu}}^{\alpha_1 \dots \alpha_{n-1}}$ and the field $G$ is defined by differentiation of the Lorentz invariant Lagrangian with respect to the field strength as
\begin{equation} \label{G} G_{\mu_1 \dots \mu_{n}}=-n!\, \frac{\partial L(F)}{\partial F^{\mu_1 \dots \mu_n}}. \end{equation}

Henceforth, we employ the following $T\bar{T}$ operator to deform the $D\!=\!2n$ dimensional electrodynamic theory\footnote{In our convention the $T\bar{T}$ operator is equal to $\frac{1}{8}\, {O}_{T^2}$ which corresponds to the definition in ref.~\cite{Ferko:2019oyv}.}
\begin{equation}
\label{TT3}
{O}_{T^2}^{[r={1}/{n}]}=T_{\mu\nu}T^{\mu\nu}-\frac{1}{n} {T_{\mu}}^{\mu} {T_{\nu}}^{\nu},
\end{equation}
which is consistent with deformations in $D\!=\!2$ and $D\!=\!4$ dimensions given by \eqref{TT1} for $n=1$ and $n=2$, respectively. This paper is an attempt to construct the deformed action from a free Lagrangian with a simple integration technique, perturbatively. In fact, $L_{free}$ denotes the Lagrangian for the local fields and the coupling constant, denoted by $\lambda$, that controls the strength of interaction among the fields as well as with local sources.

As starting point for our discussion, we compute the EM \eqref{EMT1} for a $L_{free}$ and then construct the $T\bar{T}$ operator ${O}_{T^2}={O}_{0}$ at the order of $\lambda^0$. According to the relation ${O}_{T^2}=8 \frac{\partial L_{\lambda}}{\partial \lambda}$ in ref.~\cite{Ferko:2019oyv}, the integration of this operator yields the deformed Lagrangian to the first order of $\lambda$
\begin{equation}
\label{L1}
L'_{\lambda}=L_{free}+L_1=L_{free}+\frac18 \int O_0 \,d\lambda.
\end{equation}
Using $L'_{\lambda}$ and \eqref{TT3}, one can obtain the Lagrangian of order $\lambda^2$ as
\begin{equation}
\label{L2}
L''_{\lambda}=L_{free}+L_1+L_2=L_{free}+\frac18 \int O_0 \,d\lambda+\frac18 \int O_1 \,d\lambda,
\end{equation}
where ${O}_{T^2}={O}_{0}+O_1$. By iterating this method, we will deform the free theory to higher orders of $\lambda$. When the trace of the EM tensor for $L_{free}$ is zero, the contribution of the second term in \eqref{TT3} would be of order $\lambda^2$, so it deforms the Lagrangian from the third order of $\lambda$. For instance, in deforming the non-linear electrodynamic theory by $T\bar{T}$ operator, the contribution of the second term will appear at the order $\mathcal{O}(F^8)$ in the action.

Since the $T\bar{T}$ operator given in \eqref{TT3} is proportional to quadratic power of EM tensor, it is expected that the symmetries of $T_{\mu\nu}$ of the non-linear electrodynamic BI theory be inherited to $O_{T^2}$. In particular, the EM tensor is invariant under electromagnetic duality ($S$-duality) in $D\!=\!4$. Moreover, as shown in ref.~\cite{Gibbons:1995ap}, this duality can be generalized to an $SL(2,R)$ symmetry, that is, $T_{\mu\nu}$ is invariant under $SL(2,R)$ transformation. An $SL(2,R)$ invariant form for $T_{\mu\nu}$ of non-linear electrodynamic theories is given in ref.~\cite{BabaeiVelni:2016qea}. This fact is the second motivation that encourages us to investigate this symmetry for $O_{T^2}$, i.e., we search for an $SL(2,R)$ invariant form of $T\bar{T}$ operator.

It has been shown in refs.~\cite{Aschieri:2008ns,Aschieri:2013nda} that for each non-linear electrodynamic action, as a function of some Lorentz invariant variables, the variation of the action with respect to an $SL(2,R)$ invariant parameter is also invariant under this transformation, i.e.,
\begin{equation}
\label{inv1}
 \frac{\partial S}{\partial \lambda}=-\frac{1}{\lambda}\left(S+\frac14 \int d^4x\, G F\right)=-\frac{1}{\lambda} S_{inv},
\end{equation}
where $\lambda$ is a dimensionful parameter typically presents in a non-linear theory and all indices in the product $G F$ are contracted. Here, $ S_{inv}$ is the invariant action under $SL(2,R)$ symmetry \cite{Gaillard:1981rj}. Thus from ${O}_{T^2}=8 \frac{\partial L_{\lambda}}{\partial \lambda}$ we have
\begin{equation}
\label{inv2}
 \mathcal{O}_{T^2}=\int d^4 x \,{O}_{T^2}=-\frac{8}{\lambda} S_{inv},
\end{equation}
which implies the $T\bar{T}$ operator inherits $SL(2,R)$ symmetry from the EM tensor. We show that eq.~\eqref{inv2} works for non-linear electrodynamic theories at any order of coupling expansion in $D\!=\!4$ and also for $D\!=\!2n$ with $r=1/n$ in eq.~\eqref{TT1}.

The non-linear electrodynamic theories emerge as the generalization of Maxwell theory that depend only on fields $F$ not their derivatives. In fact, it was demonstrated that the Maxwell Lagrangian is the leading-order term in the expansion in $F$ of nonlinear electrodynamic theories. These actions satisfy the Noether-Gaillard-Zumino (NGZ) identity \cite{Gaillard:1981rj}, however they are not invariant under the $S$-duality transformation while the equations of motion and EM tensor are \cite{Green:1996qg}. In four dimensions, we consider the $S$-duality transformations in extended version of the Maxwell theory in the presence of background fields, and then find the corresponding $T\bar{T}$ operators in different orders of coupling constant. Also we reconstruct the $T\bar{T}$ operator in a manifestly $SL(2,R)$ invariant form.

It has been shown in refs.~\cite{Green:1996qg,BabaeiVelni:2016qea} that the EM tensor of a non-linear theory in which dilaton and axion fields are added to BI theory, is invariant under the $SL(2,R)$ transformation ($S$-duality relevant to string theory) at any order in $\alpha'$. It was shown that the EM tensor of such theories is constructed from $SL(2,R)$ structures at any order in $\alpha'$ \cite{BabaeiVelni:2016qea}.

The other motivation about the implication of $T\bar{T}$ deformation that could be overlooked is how it affects a supersymmetric theory. Although several attempts have been made in refs.~\cite{Baggio:2018rpv,Chang:2018dge,Jiang:2019hux} for $T\bar{T}$ deformation in 2D supersymmetric theories, but it would be of interest to investigate this concept in higher dimensions. Specially, this proposal has been discussed in refs.~\cite{Chang:2019kiu,Ferko:2019oyv} for non-linear BI supersymmetric theories, $\mathcal{N}=(2, 2)$ models in $D\!=\!2$ as well as $\mathcal{N}=1$ models in $D\!=\!4$ dimensions. In this paper we will study the $T\bar{T}$ operator of  $\mathcal{N}=2$ supersymmetric theory with $U(1)$-type duality in $D\!=\!4$ that has been discussed in refs.~\cite{Kuzenko:2000uh,Broedel:2012gf}.

The structure of this paper is organized as follows: in section \ref{sec2}, we review the structure of $T\bar{T}$ operator for a chiral boson in $D\!=\!2$ and show that the deformation of free theory yields a BI-type expression which exactly agrees with the solution of flow equation obtained in different texts. In section \ref{sec3}, we investigate the $T\bar{T}$ deformation for a general non-linear BI theory of electrodynamics in arbitrary dimension $D\!=\!2n$ for even $n$. We show that the deformed Lagrangian constructed from $T\bar{T}$ operator in $D\!=\!4$ is compatible with the expansion of non-linear BI theories while it is inconsistent with related theory for $D\geq8$. In section \ref{sec4}, we investigate the $T\bar{T}$ deformation for non-linear extensions of BI theory in a non-supersymmetric model as well as $\mathcal N=2$ supersymmetric theory. In fact, we will find $T\bar{T}$ operators which are invariant under $SL(2,R)$ duality symmetry. Finally, the section \ref{sec5} is devoted to giving a brief summary of results and concluding with some outlook.
\section{$T\bar{T}$ deformation in $D=2$}
\label{sec2}

At the forefront of current research we shall consider the structure of $T\bar{T}$ deformation for a 2D scalar field theory and show that how our proposal, described by the relations \eqref{TT3}-\eqref{L2} with $n=1$, works in $D=2$. The results agree with the expression obtained in refs.~\cite{Smirnov:2016lqw,Cavaglia:2016oda,Bonelli:2018kik} which gives some confidence that the deformed theory is well defined. Suppose that $\Phi(x)$ is a dynamical 0-form scalar field (or a chiral boson) whose field strength is the vector field $F_{\mu}=\partial_{\mu} \Phi(x)$. Henceforth, for convenience we express function $\Phi(x)$ as the abbreviated form without functionality of $x$. The initial Lagrangian in undeformed theory is given by the free Lagrangian $L_{free}=-\frac12 \partial_{\alpha}\Phi \, \partial^{\alpha} \Phi$, such that we expect the deformed Lagrangian to depend on the fields only through $\partial_{\alpha}\Phi \, \partial_{\beta} \Phi$.

On the other hand, because of diffeomorphism invariance we also expect that the deformed Lagrangian is only a function of two scalar variables $\lambda$ and $\partial_{\alpha}\Phi \, \partial^{\alpha} \Phi$, i.e. $L(\lambda, P_1)$ where for later convenience we have defined $P_1=\frac12 \partial_{\alpha}\Phi \, \partial^{\alpha} \Phi$. Using the $T\bar{T}$ operator in eq. \eqref{TT3} with $n=1$ and the free Lagrangian, we can deform this theory to first order of $\lambda$ and so on. In this respect, from eqs.~\eqref{G} and \eqref{EMT1} one finds that  $G_{\mu}(\lambda^{0})=\partial_{\mu}\Phi$ and the EM tensor is
\begin{equation}
\label{EMT20}
T_{\mu\nu}(\lambda^{0})=-\tfrac12 g_{\mu\nu} \partial_{\alpha}\Phi \, \partial^{\alpha} \Phi+\partial_{\mu}\Phi\,\partial_{\nu}\Phi,
\end{equation}
which can be easily verified that it is traceless and therefore, from \eqref{TT3} we obtain
\begin{equation}
O_{T^2}(\lambda^0) =O_{0} = \tfrac{1}{2} (\partial_{\alpha}\Phi \, \partial^{\alpha} \Phi)^2.
\end{equation}
 Substituting this operator in \eqref{L1}, the deformed Lagrangian of 2D QFT up to order $\lambda$ is
\begin{equation}
\label{dL1}
L'_\lambda=-\tfrac12 \partial_{\alpha}\Phi \, \partial^{\alpha} \Phi+\tfrac{1}{16}\lambda\, (\partial_{\alpha}\Phi \, \partial^{\alpha} \Phi)^2.
\end{equation}

Now, following a similar prescription we deform the Lagrangian to the next order of $\lambda$, that is again by starting from Lagrangian \eqref{dL1}, the $G$ tensor simply is
\begin{equation}
\label{G21}
 G_\mu(\lambda )=\left(1- \tfrac{1}{4} \lambda \,\partial_{\alpha}\Phi\, \partial^{\alpha}\Phi\right) \partial_{\mu}\Phi,
\end{equation}
and explicitly the EM tensor in this order becomes
\begin{equation}
\label{EMT21}
T_{\mu \nu}(\lambda) = \left(- \tfrac{1}{2} \partial_{\alpha}\Phi \,\partial^{\alpha}\Phi + \tfrac{1}{16}\lambda (\partial_{\alpha}\Phi \, \partial^{\alpha} \Phi)^2\right) \mathit{g}_{\mu \nu} +\left(1- \tfrac{1}{4} \lambda \,\partial_{\alpha}\Phi\, \partial^{\alpha}\Phi\right)\,\partial_{\mu}\Phi \,\partial_{\nu}\Phi\,.
\end{equation}
Thus, the contribution of \eqref{dL1} to $T\bar{T}$ operator in the form $O_{T^2}(\lambda)=O_{0}+O_{1}$ is as follows
\begin{equation}
O_{T^2}(\lambda)= \tfrac{1}{2} (\partial_{\alpha}\Phi \, \partial^{\alpha} \Phi)^2- \tfrac{1}{4}\lambda \, (\partial_{\alpha}\Phi \, \partial^{\alpha} \Phi)^3,
\end{equation}
where according to relation \eqref{L2} this operator provides for us the Lagrangian of order $\lambda^2$
\begin{equation}
\label{dL2}
L''_{\lambda}=- \tfrac{1}{2} \partial_{\alpha}\Phi \,\partial^{\alpha}\Phi + \tfrac{1}{16} \lambda (\partial_{\alpha}\Phi \, \partial^{\alpha} \Phi)^2- \tfrac{1}{64}\lambda^2  (\partial_{\alpha}\Phi \, \partial^{\alpha} \Phi)^3.
\end{equation}

Iterating these steps respectively, the EM tensor to the order $\lambda^2$ is given by
\begin{eqnarray}
\label{EMT22}
T_{\mu \nu}(\lambda^2) &=&\left ( - \tfrac{1}{2} \partial_{\alpha}\Phi \partial^{\alpha}\Phi + \tfrac{1}{16} \lambda (\partial_{\alpha}\Phi \, \partial^{\alpha} \Phi)^2- \tfrac{1}{64}\lambda^2 (\partial_{\alpha}\Phi \, \partial^{\alpha} \Phi)^3 \right) \mathit{g}_{\mu \nu}  \nonumber\\
&&\,+\, \left(1- \tfrac{1}{4} \lambda\, \partial_{\alpha}\Phi\, \partial^{\alpha}\Phi+\tfrac{3}{32} \lambda^2  (\partial_{\alpha}\Phi \, \partial^{\alpha} \Phi)^2\right)\partial_{\mu}\Phi\, \partial_{\nu}\Phi,
\end{eqnarray}
which is not traceless and have a contribution in the second term of $T\bar{T}$ operator in \eqref{TT3}. The operator to order $\lambda^2$, i.e. $O_{T^2}(\lambda^2)=O_{0}+O_{1}+O_{2}$, is given by
\begin{equation}
O_{T^2}(\lambda^2)= \tfrac{1}{2} (\partial_{\alpha}\Phi \, \partial^{\alpha} \Phi)^2- \tfrac{1}{4}\lambda  (\partial_{\alpha}\Phi \, \partial^{\alpha} \Phi)^3
+\tfrac{1}{256} (34 - \frac{4}{n}) \lambda^2 (\partial_{\alpha}\Phi \, \partial^{\alpha} \Phi)^4.
\end{equation}
Thus, if we set $n\!=\!1$, the deformed Lagrangian to order $\lambda^3$ is given by $L'''_{\lambda}=L_{free}+{1}/{8} \int O_{0} \,d\lambda+{1}/{8} \int O_{1}\, d\lambda+{1}/{8} \int O_{2}\, d\lambda$ as
\begin{equation}
\label{dL3}
L'''_{\lambda}=-\tfrac{1}{2} \partial_{\alpha}\Phi \,\partial^{\alpha}\Phi + \tfrac{1}{16} \lambda (\partial_{\alpha}\Phi \, \partial^{\alpha} \Phi)^2- \tfrac{1}{64}\lambda^2  (\partial_{\alpha}\Phi \, \partial^{\alpha} \Phi)^3+\tfrac{5}{1024} \lambda^3 (\partial_{\alpha}\Phi \, \partial^{\alpha} \Phi)^4.
\end{equation}

Nonetheless, after not so hard calculations we found that the resultant deformed Lagrangian to all orders is equivalent to the expansion of a BI type Lagrangian
\begin{equation}
\label{BI2}
L_{BI}=\frac{2}{\lambda} \left(1-\sqrt{1+\frac12 \lambda\,\partial_{\alpha}\Phi \, \partial^{\alpha} \Phi}\right).
\end{equation}
This Lagrangian is exactly equal to the solution of flow equation described by the deformation operator calculated in refs.~\cite{Cavaglia:2016oda,Bonelli:2018kik} but with the deformation constant $t=\lambda/4$ and overall minus sign, because their 2D free theory is described by $L_{free}=\frac12 \partial_{\alpha}\Phi \, \partial^{\alpha} \Phi$. Here, the difference in the minus sign refers to our convention for electrodynamic theories.

Due to this fact, we expect that there is a similar shift in the levels of energy in the spectrum. According to the perturbative method in \eqref{TT3}, we can determine the energy spectrum for different orders of 2D action. It has been shown in ref.~\cite{Brennan:2020dkw} that the $T\bar{T}$ deformation of 2D QFT on AdS$_2$ background is well-defined and solvable at the quantum level.

The flow equations of the energy spectrum for a free scalar field has been derived perturbatively in ref.~\cite{Brennan:2020dkw}, in analogy with the flat space case \cite{Rosenhaus:2019utc}. The result of our calculation is given by
\begin{equation}
\label{E1}
E_m=-\frac{4\pi a}{\lambda }\left(1-\sqrt{1+\frac{\lambda E_m^{(0)}}{2\pi a}}\right),
\end{equation}
where $a$ is a parameter of length dimension and $E_m^{(0)}$ is the ground state energy. Note that our notation is different with convention in ref.~\cite{Brennan:2020dkw} by a factor $\frac14$ in defining the parameter $\lambda$. In the following, we obtain an expansion for the energy spectrum given in eq.~\eqref{E1}  in the limit $\lambda=0$
\begin{equation}
\label{E2}
E_m=\sum_p \lambda^p E_m^{(p)}= E_m^{(0)}-\lambda \frac{(E_m^{(0)})^2}{8 \pi a}+\lambda^2\frac{(E_m^{(0)})^3}{32 \pi^2 a^2}+\dots\,.
\end{equation}
The deformation of energy obtained from eq.~\eqref{TT3} with $n=1$ gives some confidence that the deformed theory from our proposal is well-defined and that the contribution of term with coefficient $\frac{1}{n}$ starts from the order $\lambda^3$.

\section{$T\bar{T}$ operator in arbitrary $D=2n$}
\label{sec3}
As the main purpose of this paper, in this section we attempt to study the structure of $T\bar{T}$ operator in general $D=2n$ dimensional BI theory of self-interacting form fields when $n$ is even. For notational convenience, we introduce two Lorentz invariant variables as
\begin{equation}
\label{P1P2}
P_1=\frac12 F_{\mu_1 \mu_2 \dots \mu_n}F^{\mu_1 \mu_2 \dots \mu_n},\quad P_2=\frac{1}{16} \left(F_{\mu_1 \mu_2 \dots \mu_n}\tilde{F}^{\mu_1 \mu_2 \dots \mu_n}\right)^2,
\end{equation}
where $ F_{\mu_1 \mu_2 \dots \mu_n}$ is the totally antisymmetric field strength denoted in section \ref{sec1} and its Hodge dual field is defined in a standard manner by
\begin{equation}
\label{HD}
\tilde{F}^{\mu_1 \mu_2 \dots \mu_n}=\frac{1}{n!} \epsilon^{\mu_1 \dots \mu_n \nu_1\dots\nu_n}F_{\nu_1 \dots \nu_n},
\end{equation}
which satisfies the following duality rotation
\begin{equation}
\label{DR}
\tilde{\tilde{F}}_{\mu_1 \dots \mu_n}=-F_{\mu_1 \dots \mu_n}.
\end{equation}

The invariance under this rotation can be extended to the electromagnetic fields interacting with the gravitational field, that does not transform under duality in the Einstein frame. One can write any non-linear electrodynamic Lagrangian as a function of these local Lorentz invariants \cite{Buratti:2019cbm}. For example, the 4D BI theory of electrodynamics is described by the Lagrangian $L_{BI}=1-\sqrt{-det(\eta_{\mu\nu}+F_{\mu\nu})}$ where $\eta_{\mu\nu}$ is the Minkowski metric. At the lowest order of the expansion of $L_{BI}$ we have the free Maxwell theory denoted by $L_{free}=-\frac14 F_{\mu\nu}F^{\mu\nu}=-\frac12 P_1$. The generalization of this Lagrangian in $D\!=\!2n$ dimensions is also represented by \cite{Gaillard:1997rt,Buratti:2019cbm}
\begin{equation}
\label{BI1}
L_{BI}=\frac{2}{\lambda n!} \left(1-\sqrt{1+\lambda P_1-\lambda^2 P_2}\right),
\end{equation}
where $\lambda$ is a coupling constant or an expansion parameter such that $L_{free}=-\frac{1}{n!} P_1$ is at the order of $\lambda^0$.\footnote{We obtained a similar BI-type theory in section \ref{sec2} for 2D QFT with $P_1\!=\!\frac12 \partial_{\alpha}\Phi \, \partial^{\alpha} \Phi$ and $P_2=\frac{1}{16}\,\partial_{\mu}\Phi \,\epsilon^{\mu\alpha}\, \partial_{\alpha}\Phi \,  \partial_{\nu}\Phi \,\epsilon^{\nu\beta}\,\partial_{\beta} \Phi,$ where $\epsilon^{\mu\nu}$ is the antisymmetric Levi-Civita symbol. Therefore, due to the symmetry of partial derivatives $P_2=0$ and the BI Lagrangian is $L_{BI}=\frac{2}{\lambda} \left(1-\sqrt{1+ \lambda\,P_1}\right)$ for $n=1$ which is compatible with \eqref{BI2}.}

It has been shown in ref. \cite{Ferko:2019oyv} that the flow equation for the deformed Lagrangian is proportional to $T\bar{T}$ operator in two dimensions. In order to investigate this possibility in higher dimensional non-linear BI theory, we use the definition of EM tensor given in \eqref{EMT1}. Thus, we obtain
\begin{equation}
\label{TT4}
T_{\mu\nu}T^{\mu\nu}=D L^2+\frac{2}{(n-1)!}\, (F^{\mu}G_{\mu} )\, L+\frac{1}{[(n-1)!]^2}(F^{\mu}G^{\nu} ) (F_{\mu}G_{\nu} ).
\end{equation}
According to \eqref{BI1} the Lagrangian can be defined as a function of variables $P_1$ and $P_2$, and the coupling constant $\lambda$, i.e. $L_{BI}(P_1,P_2,\lambda)$, such that its expansion in the zero coupling limit becomes
\begin{equation}
\label{exp1}
L_{BI}\!=\!\frac{1}{n!}\!\left[-P_1\!+\!\frac{\lambda}{4}(P_1^2\!+\!4P_2)\!-\!\frac{\lambda^2}{8} P_1(P_1^2\!+\!4P_2)\!+\!\frac{\lambda^3}{64}(5P_1^2\!+\!4P_2)(P_1^2\!+\!4P_2)\!+\!\mathcal{O}(F^{10})\right],
\end{equation}
therefore, to disclose the relation between $O_{T^2}$ and $\frac{\partial L_{\lambda}}{\partial \lambda}$ both ${T_{\mu}}^{\mu}$ and $T_{\mu\nu}T^{\mu\nu}$ should be functions of these parameters. From \eqref{EMT1}, the trace of the EM tensor includes a term $(F_{\mu}G^{\mu})$ beside the Lagrangian which is also a function of $P_1$, $P_2$, and $\lambda$. So from this expression one finds that ${T_{\mu}}^{\mu}$ satisfies the following equation
\begin{equation}
\label{TEM}
\frac{1}{D}{T_{\mu}}^{\mu}=-\lambda \frac{\partial L_{\lambda}}{\partial \lambda}.
\end{equation}

On the other hand, the first term in $O_{T^2}$, i.e. $T_{\mu\nu}T^{\mu\nu}$, have a term described by a tensor $K_2=(F^{\mu}G^{\nu})(F_{\mu}G_{\nu})$. The most striking feature which distinguishes our calculation for lower dimensional studies in $T\bar{T}$ deformation, is the presence of tensor $K_2$ in this operator. It can be verified that $K_2$ is written exactly as a function of $P_1$ and $P_2$ in $D\!=\!2$ and $D\!=\!4$ but in higher dimensions $D>4$ this does not happen. We postpone this discussion to the next subsections. In the context of duality invariant formalism in \cite{Gibbons:1995ap,Gaillard:1997rt}, more details about the most general Lorentz invariant Lagrangian can be found in refs.~\cite{Buratti:2019cbm,Buratti:2019guq}.

In the rest of this section we consider the proposal
\begin{equation} \label{prop}
{O}_{T^2}\equiv T_{\mu\nu}T^{\mu\nu}-\frac{1}{n} {T_{\mu}}^{\mu} {T_{\nu}}^{\nu}= 8 \frac{\partial L_{\lambda}}{\partial \lambda},
\end{equation}
in $D\geq 4$ spacetime dimensions with more details. We show that the coefficient $r$ in four dimensions is equal to 1/2, however in $D>4$, the tensor $K_2$ precludes this possibility. We will also obtain a general form of deformed Lagrangian in arbitrary $D\!=\!2n$ dimensions at each order of $\lambda$.
\subsection{$T\bar{T}$ deformation in $D\!=\!4$}
\label{sec31}

In 4D where $n=2$, we have a 2-form field strength $F_{\mu\nu}$ and the Lorentz invariant variables are
\begin{equation}
\label{4dP1P2}
P_1= \frac12  F_{\mu\nu} F^{\mu\nu},\quad P_2=\frac{1}{16} \left(F_{\mu\nu} \tilde{F}^{\mu\nu}\right)^2.
\end{equation}
Using the expansion in \eqref{exp1} for $\lambda$ and the definition \eqref{G}, we compute the tensor $G_{\mu\nu}$ as
\begin{eqnarray}\label{G3}
G_{\mu \nu}&=&F_{\mu \nu} - \lambda (\tfrac{1}{8}  \tilde{F}_{\mu \nu} F_{\alpha \beta}\tilde{F}^{\alpha \beta} +  \tfrac{1}{2} F_{\mu \nu} P_1) + \lambda^2 \bigl(\tfrac{1}{16} \tilde{F}_{\mu \nu} F_{\alpha \beta}\tilde{F}^{\alpha \beta} P_1 + \tfrac{1}{8} F_{\mu \nu} (3 P_1^2 + 4 P_2)\bigr) \nonumber\\
&-&   \lambda^3 \bigl( \tfrac{1}{64} \tilde{F}_{\mu \nu} F_{\alpha \beta}\tilde{F}^{\alpha \beta} (3 P_1^2 + 4 P_2) +  \tfrac{1}{16} F_{\mu \nu} P_1 (5 P_1^2 + 12 P_2)\bigr)+...\,,
\end{eqnarray}
then, by substituting this result in eq.~\eqref{EMT1}, the EM tensor becomes
\begin{eqnarray}
\label{EMT4}
T_{\mu \nu} &=&F_{\mu}{}^{\alpha} F_{\nu \alpha} -  \tfrac{1}{2} P_1 \mathit{g}_{\mu \nu} - \lambda ( \tfrac{1}{2} F_{\mu}{}^{\alpha} F_{\nu \alpha} P_1 - \tfrac{1}{8} P_1^2 \mathit{g}_{\mu \nu}) \nonumber\\
&+& \lambda^2 \bigl(\tfrac{1}{8} F_{\mu}{}^{\alpha} F_{\nu \alpha} (3 P_1^2 + 4 P_2) -  \tfrac{1}{16} P_1^3 \mathit{g}_{\mu \nu}\bigr)  \\
&-& \lambda^3 \bigl(\tfrac{1}{16} F_{\mu}{}^{\alpha} F_{\nu\alpha} P_1 (5 P_1^2 + 12 P_2) - (\tfrac{5}{128} P_1^4 -  \tfrac{1}{8} P_2^2) \mathit{g}_{\mu \nu}\bigr)+...\,,\nonumber
\end{eqnarray}
where the elliptic terms in both equations represent higher order powers of $\lambda$ expansion. We have also checked that the trace of \eqref{EMT4} is compatible with \eqref{TEM} for $D\!=\!4$.

The tensor $K_2$ in 4D is written in the form
\begin{equation}
\label{k2}
K_2=F^{\mu\alpha}{G^{\nu}}_{\alpha}\,F_{\mu\beta}{G_{\nu}}^{\beta},
\end{equation}
where from the relation \eqref{G3} we can rewrite it as
\begin{eqnarray}\label{k22}
K_2 &=&K_0 - \lambda ( K_0 P_1 + 2 P_1 P_2) + \lambda^2 \bigl(K_0 (P_1^2 + P_2) +  P_2 (2 P_1^2 +  P_2)\bigr)  \nonumber\\
&-& \lambda^3 \Bigl( K_0 P_1 (P_1^2 + 2 P_2) +   P_1 P_2 \bigl(2 P_1^2 + 3 P_2\bigr)\Bigr)+...\,,
\end{eqnarray}
where $K_0=F^{\mu\alpha}{F^{\nu}}_{\alpha}\,F_{\mu\beta}{F_{\nu}}^{\beta}$.
If we attempt to redefine $K_0$ in terms of Lorentz variables $P_1$ and $P_2$ then $T_{\mu\nu}T^{\mu\nu}$ can be expressed only as a function of $P_1, P_2$, and $\lambda$. It can be written in implicit form $K_0=2(P_1^2+2P_2)$. From this expression it is clear that we can test our suggestion \eqref{prop} and determine the coefficient $r$ for each order of $\lambda$ expansion. The result of calculations has forcefully stressed that $r=1/2$ which is compatible with the results obtained in ref.~\cite{Conti:2018jho}.

Following the perturbative method mentioned in section \ref{sec1} by eqs.~\eqref{TT3} and \eqref{L1}, we can deform the Maxwell action at each order of $\lambda$ in 4D and give the expansion of Maxwell-BI theory. This is reminiscent of the fact that we can expect a similar deformation for the energy spectrum. Thus, one can find a solvable energy spectrum in $D\!=\!4$ according to \eqref{exp1}. In other words, applying the $T\bar{T}$ deformation to the energy spectrum of the Maxwell action yields the one for BI theory which in spite of Maxwell theory, its energy spectrum is finite and solvable. It has been shown in ref.~\cite{Rasheed:1997ns} that for purely electric configuration in flat space, the total self-energy of the point charge is finite and integrable. For the Lagrangian
\begin{equation}
L_{BI} = {1\over \lambda} \left\{ 1 - \sqrt{1-\lambda \mbox{\boldmath E}^2}
\right\},
\end{equation}
there is an upper bound on the electric field strength $\boldmath E$
\begin{equation}
\left|\mbox{\boldmath E}\right| \le {1\over \sqrt{\lambda}}.
\end{equation}
The total self-energy of the point charge with $E_r = {Q\over\sqrt{r^4+\lambda Q^2}}$  is
\begin{equation}
{\mathcal E} = {1\over 16\pi} \int d^3 x T_{00} = {1\over
  16\pi} \int d^3 x {1\over
  \lambda r^2}\left(\sqrt{r^4+\lambda Q^2}-r^2\right).
\end{equation}
Integrating this energy yields
\begin{equation}
{\cal E} = {2Q^2\over 12} \int_0^\infty {dr\over\sqrt{r^4+\lambda Q^2}} =
{(\pi Q)^{3\over 2}\over 12\sqrt{\lambda^{1/2}}\,\Gamma\!\left({3\over 4}\right)^2}.
\end{equation}
Therefore, the BI theory of electrodynamics succeeded in its original goal of providing a model for point charges with finite self-energy. Note that in the limit $\lambda\rightarrow 0$, Maxwell theory is reproduced and the self-energy diverges.
\subsection{$T\bar{T}$ deformation in $D=8$}
\label{sec32}

Following the general strategy used in the previous subsection, we define two Lorentz invariant parameters $P_1$ and $P_2$ in terms of 4-form field strengths as follows
\begin{equation}
\label{8dP1P2}
P_1=\frac12 F_{ \mu\nu\rho \sigma} F^{\mu\nu\rho \sigma},\quad P_2=\frac{1}{16}\left(F_{\mu\nu \rho \sigma} {\tilde{F}}^{\mu\nu \rho \sigma}\right)^2,
\end{equation}
where from the expansion \eqref{exp1} and definition \eqref{G}, the $G$ tensor is given by
\begin{equation}
\label{G8}
 G_{\mu\nu\rho \sigma}=\frac{1}{12} F_{\mu\nu\rho \sigma} - \tfrac{1}{96} \lambda (4 F_{\mu\nu\rho \sigma} P_1 +  \tilde{F}_{\mu\nu\rho \sigma} F^{\alpha \beta \epsilon \delta} \tilde{F}_{\alpha \beta \epsilon \delta})+...\,.
\end{equation}
Substituting the above results in eq.~\eqref{EMT1} yields the deformed EM tensor of 8D electrodynamic theory as
\begin{eqnarray}
\label{EMT8}
T_{ \mu \nu}&=&\tfrac{1}{6} F_{\mu}{}^{ \alpha\beta \gamma} F_{\nu \alpha\beta \gamma} -  \tfrac{1}{24} P_1 \mathit{g}_{\mu \nu} \nonumber\\
&+& \lambda \biggl(\tfrac{1}{384} (4 P_1^2 + 16 P_2) \mathit{g}_{\mu \nu} - \tfrac{1}{48} (4 F_{\mu}{}^{\alpha\beta \gamma} F_{\nu \alpha\beta \gamma} P_1 +  F^{\eta \rho \sigma \delta} {\tilde{F}}_{\eta \rho \sigma \delta} {\tilde{F}}_{\mu \alpha\beta \gamma} F_{\nu}{}^{\alpha\beta \gamma})\biggr)\nonumber\\&+&\mathcal{O}(\lambda^2).
\end{eqnarray}

Without a doubt it can be found that the trace of this tensor satisfies the condition \eqref{TEM} in $D\!=\!8$ dimensions.
In this case the tensor $K_2$ in the last term of eq.~\eqref{TT4} can be written as
\begin{equation}
\label{K24} K_2 =\tfrac{1}{36} K_0 -  \tfrac{1}{288} \lambda P_1 (8 K_0+ 3 P_1^2 + 8 P_2)+...\,,
\end{equation}
where $K_0= F^{\mu \alpha\beta \gamma}{F^{\nu}}_{\alpha\beta \gamma}\,\,F_{\mu\rho\sigma\delta}{F_{\nu}}^{\rho\sigma\delta}$. Thus, the quadratic power of EM tensor becomes
\begin{equation}
\label{TT8}
T_{\mu \nu} T^{\mu \nu}=\frac{1}{72} (2 K_0 -P_1^2)-\frac{1}{72}  \lambda  (2 K_0 P_1-P_1^3)+...\,.
\end{equation}

In order to be able to express the covariant variables $T_{\mu \nu} T^{\mu \nu}$ and $(T_{\mu}{}^{ \mu})^2$ as a function of Lorentz variables we should write $K_0$ in terms of these parameters, however in spite of four dimensionl theory, it is not possible to do this in $D=8$. This possibility has been considered in refs. \cite{Buratti:2019cbm,Buratti:2019guq} and it is shown that $K_0$ recasts as
\begin{equation}
\label{k0}
K_0=\frac{1}{4} P_1^2+\frac{1}{2} P_2 +\frac{9}{8} K_3,
\end{equation}
where $K_3=(F^{\mu\nu }{F^{ \alpha \beta}})\,\,(F_{\mu\nu}{F_{\alpha \beta}})=F^{\mu\nu \gamma \eta}{F^{ \alpha \beta}}_{ \gamma \eta}\,\,F_{\mu\nu\epsilon\delta}{F_{\alpha \beta}}^{\epsilon \delta}$ is an independent tensor in $D=8$ \cite{Buratti:2019cbm}.  Therefore, we are unable to compare the variation of the Lagrangian \eqref{BI1} with respect to $\lambda$, which only depends on $P_1$ and $P_2$, with $O_{T^2}$ operator through eq.~\eqref{prop}. In this respect, we cannot write the variation of 8D BI action in terms of $O_{T^2}$ like in \eqref{inv2} (up to $\mathcal {O}(F^4)$ and $\mathcal {O}(F^6)$).

Here, we apply the prescription explained in section \ref{sec1} to deform the Lagrangian in favor of $T\bar{T}$ operator. Starting from the free Lagrangian in $D=8$, i.e.  $L_{free}=-\frac{1}{4!} P_1=-\frac{1}{2\times 4!} F_{\mu\nu\alpha\beta} F^{\mu\nu\alpha\beta}$, and using \eqref{TT8}, we can deform the free Lagrangian to the first order in $\lambda$ with operator $O_{T^2}=O_{0}=\frac{1}{72} (2 K_0 -P_1^2)$ as the following form
\begin{equation}
\label{defL1}
L^{\prime}_{\lambda}=L_{free}+\frac{1}{8} \int O_{0} d\lambda=-\frac{1}{4!} P_1+\frac{1}{576} \lambda (2 K_0 -  P_1^2).
\end{equation}
Now, one can employ this Lagrangian and define the EM tensor to obtain $O_{T^2}(\lambda)=O_{0}+O_{1}$ in which
\begin{eqnarray}
\label{TTo1}
O_{1}&=&\frac{\lambda}{72} \bigg(-\frac13 (F_{\alpha}F^{\delta})( F^\alpha F^{ \beta})( F_\beta F_{ \delta}) -( F^\alpha F^{ \beta} )(F^\delta F^{ \zeta})( F_{\beta \delta }F_{ \alpha \zeta})\nonumber\\
&&\hspace{1cm}+ \frac13 ( F_\alpha F_{ \beta})( F^\alpha F^{ \beta})( FF) -\frac{1}{48} (FF)^3 \bigg),
\end{eqnarray}
where all pairs are contractions of 4-form field strengths. If we iterate this procedure the next order in $\lambda$ is constructed. So from \eqref{L2}, the Lagrangian upto $\mathcal{O}(F^6)$ deforms as
\begin{eqnarray}
\label{L20}
L''_{\lambda}&=&- \frac{1}{48} (FF) + \frac{\lambda}{288}  \bigg(( F_\alpha F_{ \beta})( F^\alpha F^{ \beta}) -\frac18  (FF)^2\bigg)
- \tfrac{\lambda^2}{1152}  \bigg(\frac13 (F_{\alpha}F^{\delta})( F^\alpha F^{ \beta})( F_\beta F_{ \delta})\nonumber\\
&& + ( F^\alpha F^{ \beta} )(F^\delta F^{ \zeta})( F_{\beta \delta }F_{ \alpha \zeta})- \frac13( F_\alpha F_{ \beta})( F^\alpha F^{ \beta})( FF) +\frac{1}{48}  (FF)^3 \bigg).
\end{eqnarray}

Since the Lagrangian $L''_{\lambda}$ is derived from $O_{T^2}$ operator of order $\lambda$ and the fact that the second term with $\frac{1}{n}$ coefficient in $T\bar{T}$ deformation comes from $\lambda^2$ order, this term does not contribute in the Lagrangian \eqref{L20}. Thus, we should consider the next order in $O_{T^2}$ operator to examine the contribution of this term, i.e., $O_{T^2}(\lambda^2)=O_{0}+O_{1}+O_{2}$. Due to this fact, we obtain
\begin{eqnarray}\label{O8}
O_{2}&=& \frac{1}{27648} \lambda^2 \bigg(  \tfrac{88}{3}( F_\alpha F_{ \beta})( F^\alpha F^{ \beta})( F^\epsilon F^{ \delta})( F_{ \epsilon} F_\delta ) - 24\, (F_{\alpha}F^{\delta} )(F^\alpha F^{ \beta} )(F_\beta F_{ \delta} )(FF )\nonumber  \\
&+& \tfrac{46}{3}( F_\alpha F_{\beta})( F^\alpha F^{ \beta} )(FF)^2 -  \tfrac{5}{8} (FF)^4-  \tfrac{64}{3}( F^\alpha F^{ \beta})( F_{\beta}F^{\epsilon})( F_{\epsilon}F^{\delta})( F_{\alpha}^{\gamma}{}F_{\gamma \delta} ) \\
&-& 72\, (F^\alpha F^{ \beta} )(F^\epsilon F^{ \delta} )(F_{\beta \epsilon }F_{\alpha \delta} )(FF)+ 144\, (F^\alpha F^{ \beta})( F^{\epsilon}F_{\beta})( F^ \delta F^{ \gamma})( F_{\gamma \alpha }F_{ \delta\epsilon})\nonumber  \\
&+& 32\,( F^\alpha F^{ \beta})( F^\epsilon F^{ \delta} )(F^\gamma F^{ \kappa})( F_{\beta \epsilon \gamma }F_{ \alpha \delta \kappa}) + 144\, (F^\alpha F^{ \beta}) (F^\epsilon F^{ \delta})( F_{\beta}{}^{\gamma}{} F_{\alpha}{}^{\kappa} )(F_{\gamma}{}_{\epsilon }F_{\kappa  \delta})\bigg) \nonumber \\
&\!\!\!\!-\!\!\!\!&\frac{1}{n} \lambda^2 \bigg( \tfrac{1}{1296}( F_\alpha F_{\beta})( F^\alpha F^{ \beta})( F^\epsilon F^{ \delta} )( F_{ \epsilon}F_\delta) -  \tfrac{1}{5184}( F_\alpha F_{\beta})( F^\alpha F^{ \beta} )(FF)^2 + \tfrac{1}{82944}( FF)^4  \bigg). \nonumber
\end{eqnarray}
Finally, using this result we are ready to generate the Lagrangian of order $\lambda^3$ by taking $n=4$, which leads to
\begin{equation}
\label{defL83}
L'''_{\lambda}=L''_{\lambda}+\frac18 \int O_2 \, d\lambda.
\end{equation}

It is worth to mention some relevant issues about the $T\bar{T}$ deformation in diverse dimension and holographic interpretation before closing this section;  
\begin{itemize}
\item The generalization to higher dimensions is analogous to the computations done for 4D and 8D theories. We find a general form of the deformed Lagrangian at each order of deformation parameter in $D\!=\!2n$-dimensional theory of electrodynamic fields. For instance, the deformation at the order of $\lambda$ is given by
\begin{equation}
\label{gen1} L_{\lambda}\sim \lambda\left((F_{\alpha}F_{\beta})(F^{\alpha}F^{\beta})-\frac{1}{D} (FF)^2\right),
\end{equation}
and in the next order has the form
\begin{eqnarray} \label{gen2}
L_{\lambda}&\sim& \lambda^2 \bigg(-2 ( F_{\alpha} F^{\gamma })( F^\alpha F^{ \beta})( F_{\beta}F_{\gamma} )- (D-2) (F_{\alpha \gamma} F_{\beta \eta })( F^\alpha F^{ \beta}) ( F^\gamma F^{ \eta})\nonumber\\
&&+ 2 ( F_\alpha F_{ \beta})( F^\alpha F^{ \beta}) ( FF) -  \tfrac{1}{D}  (FF)^3\bigg),
\end{eqnarray}
where all the field strengths are $n$-form fields which their indices are contracted appropriately and the equality is obtained by a numerical coefficient as a function of spacetime dimension.
\item From the holography point of view, it has been shown \cite{Henningson:1998gx} that the anomaly of conformal closed algebra of the CFT is compatible with the $\mathcal{A}$ anomaly of the  $D\!+\!1$-dimensional bulk theory under a conformal transformation. For example, in the case of $D\!=\!2$ the finite term of the regularized action is given by $\mathcal{A}=-\frac{c}{24\pi}R$ with $c=3\ell/2G$ which agrees with the value of the conformal anomaly $c$ as computed in \cite{Brown:1986nw} by considering the asymptotic symmetry algebra of $AdS_{3}$ space, or in $D\!=\!4$ the anomaly $\mathcal{A}=-\frac{N^2}{\pi^2}(E_4+I_4)$ is compatible with the $\mathcal{N}=4$ superconformal $SU(N)$ gauge theory with $a=c=\frac14(N^2-1)$ in the large-$N$ limit. Similarly in $D\!=\!6$ the anomaly is proportional to $N^3$ where $N$ is the number of coincident $M5$ branes in ($0, 2$) superconformal theory \cite{Bastianelli:2000hi}. 
\\
Obviously the field content of QFTs are not the same and there are different matter fields in each sector. However, the $T\bar{T}$ operators obtained in this section are only related to the non-linear BI theories of $p$-form fields that may live on a single brane and are not necessary a CFT with a closed conformal algebra of central charges $a=c$, so it seems that there is no justification to compare our results to that derived from AdS/CFT in \eqref{TT1}.
\item Since the duality-symmetric BI theory in $D\!=\!4$ and its chiral 2-form counterpart in $D\!=\!6$ are related by dimensional reduction of the latter, one can conclude that the $T\bar{T}$ deformation of the free chiral 2-form theory is the same as in $D\!=\!4$ for the BI case. In other words, for the 6D chiral 2-form theories there is exactly the same number "two" of independent Lorentz-invariants as in $D\!=\!4$ (e.g. like $P_1$ and $P_2$ in this paper) which can be used to construct their consistent non-linear generalizations, which are always related to those in 4D theory \cite{Bandos:2020hgy}.
\\
For the non-chiral 2-form field, the trace of the EM tensor is non-zero, while for the chiral (self-dual) field it is zero. By the way, this indicates that the free chiral 2-form theory is conformal \cite{Bandos:2020jsw}. However, if we try to get a non-linear chiral p-form theory upon deformation, we should give a prescription of how the self-duality condition gets deformed under $T\bar{T}$ operator. 
\\
On the other hand, chiral theories require auxiliary fields in their actions only to ensure manifest Lorentz-invariance. Without the use of the auxiliary fields the chiral theory is still Lorentz-invariant, but not manifestly. For example, in the  $T\bar{T}$ deformation related to the M5-brane, one should start from a fully-fledged action for the free chiral 2-form and compare it with the fully-fledged action for the M5-brane. Such actions have a single scalar auxiliary field to ensure manifest Lorentz-invariance. But this field can be gauged away directly in the action. Then the resulting action will be non-manifestly Lorentz invariant but will contain only the physical 2-form field whose equation of motion produces the self-duality condition for its field strength and the required traceless EM tensor. 
 \end{itemize}  
\section{$T\bar{T}$ operator in BI type theories and S-duality}
\label{sec4}
As discussed in the previous section, according to theorems in refs.~\cite{Gaillard:1997rt,Aschieri:2008ns}, the non-linear electrodynamic theories satisfying the NGZ identity are called $S$-dual which are invariant under electromagnetic transformation at the level of the equations of motion, the EM tensor and the S-matrix. This $S$-dual symmetry can be enhanced to invariance under $SL(2,R)$ transformation. It is shown that this is true for any non-linear theory of electrodynamics. The BI theories like S-dual electrodynamic theories, are a particular class of solutions in these theorems which have exact closed forms \cite{Rasheed:1997ns}. In this section we will consider the $T\bar{T}$ deformation for non-supersymmetric BI-type theory, as well as $\mathcal{N}=2$ supersymmetric model.

\subsection{Non-supersymmetric model in $D\!=\!4$}
\label{sec41}
The Lagrangian of an electrodynamic BI theory in the presence of a dilatonic field in Einstein frame is given by the following density
\begin{equation}
\label{DL1} L_{BI} ={\lambda}^{-1} \bigg[1-\sqrt{-\det(\eta_{\mu\nu}+\lambda^{\frac{1}{2}}e^{-\phi_0/2}F_{\mu\nu})}\,\bigg],
\end{equation}
where the dilaton function $e^{\phi_0/2}$ is regarded as an effective gauge coupling constant. The expansion of this Lagrangian in the limit $\lambda=0$ becomes\footnote{Our convention here for $Tr[F^2]$ is $F_{\mu\nu}F^{\nu\mu}$ and similarly for higher powers of $F_{\mu\nu}$.}
\begin{eqnarray}\label{exp2}
L_{BI} &=&\frac{e^{-\phi_0}}{4}Tr[F^2]+\lambda \frac{e^{-2\phi_0}}{8}\bigg(Tr[F^4]-\frac{1}{4}Tr[F^2]^2\bigg)\\
&+&\lambda^2\frac{e^{-3\phi_0}}{12}\bigg(Tr[F^6]-\frac{3}{8}Tr[F^2]Tr[F^4]+\frac{1}{32}Tr[F^2]^3\bigg)+\dots\,.\nonumber
\end{eqnarray}
After computing the antisymmetric tensor $G$ from \eqref{G}, one can find the EM tensor as
\begin{eqnarray}
\label{EMTns}
T_{\mu \nu}&=& e^{\Phi_0}\bigg(\frac{1}{4}  \mathit{g}_{\mu \nu} Tr[F^2]- (F^2)_{\mu\nu} \bigg)
+ \lambda e^{- 2\phi_0 }\bigg(\frac{1}{4} (F^2)_{\mu\nu} Tr[F^2] -(F^4)_{\mu\nu} - \frac{1}{32}  \mathit{g}_{\mu \nu} Tr[F^2]\nonumber \\
&+&  \frac{1}{8} \mathit{g}_{\mu \nu} Tr[F^4] \bigg)+ \lambda^2 e^{-3\phi_0 }\bigg( \frac{1}{4}  (F^4)_{\mu\nu} Tr[F^2]-  \frac{1}{32} (F^2)_{\mu\nu} Tr[F^2]^2 + \frac{1}{8} (F^2)_{\mu \nu} Tr[F^4]\nonumber\\
& -& (F^6)_{\mu\nu}+ \frac{1}{384} \mathit{g}_{\mu \nu} Tr[F^2]^3  -  \frac{1}{32}  \mathit{g}_{\mu \nu} Tr[F^2] Tr[F^4] + \frac{1}{12}  \mathit{g}_{\mu \nu} Tr[F^6]\bigg)+\dots\,.
\end{eqnarray}

Equipped with this EM tensor, we find the $T\overline{T}$ operator from eq.~\eqref{TT1} given by
\begin{eqnarray} \label{ottns}
O_{T^2}^{[r]}&=& \frac{1}{16} e^{- 2\phi_0 }\bigg( [D-8 -  r (D-4)^2] \, Tr[F^2]^2+16\,Tr[F^4] \bigg)\nonumber\\
&+&\frac{\lambda}{64}  e^{- 3\phi_0 } \bigg([12 -  D +r(D^2-12 D+32)]\, Tr[F^2]^3\nonumber\\
&&+ 4\, [D-20 - r(D^2-12 D+32)]\, Tr[F^2] \,Tr[F^4] + 128\, Tr[F^6]\bigg)\nonumber\\
&+&\frac{\lambda^2}{3072}  e^{- 4\phi_0 } \bigg([7D-688 -r(7D^2-112D+384) ]\, Tr[F^2]^4 \nonumber\\
&&- 24\, [3D-160 +r(3 D^2 -48D+160)] \,Tr[F^2]^2 \,Tr[F^4]\nonumber\\
&&+ 48\, [D+16 -  r(D-8 )^2] \,Tr[F^4]^2\nonumber\\
&&+128\,[D-40- r(D^2-16D+16)] \,Tr[F^2] \,Tr[F^6]\bigg)+\dots\,.
\end{eqnarray}
The zeroth-order terms, i.e. $\mathcal{O}(\lambda^0)$ or $\mathcal{O}(F^4)$, is the $T\bar{T}$ operator of the free Maxwell theory. This is the same for the BI action given in \eqref{exp2} at order $\mathcal{O}(\lambda^1)$. It is obvious that the Maxwell $T\bar{T}$ operator is independent of $r$ in $D\!=\!4$. The same behaviour holds for the second-order terms of $\mathcal{O}(\lambda^1)$ in \eqref{TT1}. In fact, the $r$ dependence of $T\bar{T}$ operator of BI theory in $D\!=\!4$ starts at order $\mathcal{O}(\lambda^2)$.

On the other hand, due to the action \eqref{exp2} and the flow eqs. \eqref{inv1} and \eqref{inv2}, we can find the above $T\bar{T}$ operator in $D\!=\!4$ by setting $r=\frac12$ as follows
\begin{eqnarray}
\label{O4}
O_{T^2}^{[1/2]}&=& -\frac{1}{4}\,e^{-2 \Phi_0} \bigg( Tr[F^2]^2 -4\, Tr[F^4]\bigg)\nonumber\\
& -&  \frac{1}{8}\, e^{-3 \Phi_0} \bigg(Tr[F^2]^3 - 4 \,Tr[F^2]\, Tr[F^4]\bigg) \nonumber\\
& -&   \frac{3}{256} \,e^{-4 \Phi_0} \bigg(3\, Tr[F^2]^4 - 8\, Tr[F^2]^2\, Tr[F^4] - 16\, Tr[F^4]^2\bigg)\nonumber\\
& - &  \frac{1}{256}\, e^{-5 \Phi_0}\bigg(Tr[F^2]^5 + 8\, Tr[F^2]^3\, Tr[F^4] - 48\,  Tr[F^2]\, Tr[F^4]^2\bigg)+\dots\,,
\end{eqnarray}
where we have set $\lambda=1$ without loss of generality and used the following identity
\begin{equation}
\label{id1}
Tr[F^6]-\frac{3}{4}\,Tr[F^2]\,Tr[F^4]+\frac{1}{8}\,Tr[F^2]^3=0.
\end{equation}
\subsection{$SL(2,R)$ invariant structure of 4D $T\bar{T}$ operator}
\label{sec42}
In this subsection we review non-linear $SL(2,R)$ transformation of form fields and then reconstruct the $T\bar{T}$ operator in an $SL(2,R)$ invariant form.
In order to find the $SL(2,R)$ structure of EM tensor of a non-linear electrodynamic theory in four dimensions, we first consider the behavior of field strengths $F_{\mu \nu}$ and $G_{\mu \nu}$ under transformation
\begin{equation}
\label{sl2r}
\tau \rightarrow \frac{a\tau+b}{c\tau+d},
\end{equation}
where $\tau$ is a complex scalar field defined by $\tau=C_0+ie^{-\phi_0}$. In general, it is referred to an axion-dilaton field such that $C_0$ is an axion field and $\phi_0$ is a dilaton.

The field strengths transform as a doublet $\mathcal{F}_{\mu\nu}$ under the $SL(2,R)$ symmetry group as following \cite{Gibbons:1995ap}
\begin{equation}
\label{gf1}
\mathcal{F}_{\mu\nu}\rightarrow  (\Lambda^{-1})^T \mathcal{F}_{\mu\nu},
\end{equation}
where $\Lambda=\left(\begin{array}{cc}a&b  \\ c&d\end{array}\right)$ is an $SL(2,R)$ matrix. One can rewrite the Lagrangian in an $SL(2,R)$ invariant form by using the matrix
\begin{equation}
\label{M}
\mathcal{M}=e^{\phi_0}\left(\begin{array}{cc} |\tau|^2&-C_0  \\ -C_0&1\end{array}\right),
\end{equation}
which under S-duality transforms as ${\mathcal M}\rightarrow \Lambda\, {\mathcal M}\,\Lambda ^T$.

Now, from the above consideration we construct an $SL(2,R)$ invariant structure $ \mathcal F^T \mathcal M\mathcal F$. By separating the contribution of the axion coupling in the Lagrangian as $L=L'+C_0 z$, that yields a decomposition in $G$ as $G_{\mu \nu}=G^{'}_{\mu \nu}-C_0\tilde{F}_{\mu \nu}$, one can find the $SL(2,R)$ invariant structure in the form that the axion field does not appear, i.e.
 \begin{equation}\label{sl2r2}
(\mathcal F^T)_\mu{}^\rho\mathcal M_0\,\mathcal F_{\nu \rho}=e^{-\phi_0}\,\tilde{F}_\mu{}^\rho\,\tilde{F}_{\nu \rho}+ e^{\phi_0}\, G^{'}_{\mu}{}^\rho\, G^{'}_{\nu \rho}\,,
\end{equation}
where $z=1/4\, F^{\mu \nu} \tilde {F} _ {\mu \nu}$ is a Lorentz invariant parameter and $G'$ comes from \eqref{G} for $ L'$. Assuming that  the axion and dilaton fields are constant and do not fluctuate, henceforth we use $\mathcal M_0$ to denote this fact.

We are interested here to study the $T\bar{T}$ operator of some non-linear BI electrodynamic theories and $\mathcal N=2$ supersymmetric BI type theory. The equations of motion of these theories enjoy duality symmetries, however, the corresponding Lagrangians are not invariant under the duality transformations.

According to \eqref{EMT1}, one can find the EM tensor of the Lagrangian $L$ in terms of the field strengths $F$ and $G$. Using some identities that hold between some trace structure of $F$'s and by analogy with the $SL(2,R)$ invariant structure \eqref{sl2r2}, the EM tensor and consequently the $T\bar{T}$ operator could be appeared in terms of this $SL(2,R)$ invariant structure.

In the following, we would like to find the above operator in terms of $SL(2,R)$ invariant structures. Considering the approach in ref.~\cite{BabaeiVelni:2016qea} that was applied to find the $SL(2,R)$ invariant form of EM tensor for non-linear electrodynamic theories, the above $T\overline{T}$ operator can be found in the form that is manifestly $SL(2,R)$ invariant. We should find the non-linear $SL(2,R)$ invariant structure \eqref{sl2r2} corresponding to BI theory in terms of the field strengths $F$ and $G$. Therefore, the calculation yields
\begin{eqnarray}
\label{OT2}
O_{T^2}^{[1/2]}&=& \frac{1}{4} (\mathcal F^T)_\mu{}^\rho \mathcal M_0\,\mathcal F_{\rho\nu} (\mathcal F^T)^{\mu\sigma} \mathcal M_0\,\mathcal F_{\sigma}{}^{\nu} -  \frac{5}{64} {Tr(\mathcal F^T\mathcal M_0\,\mathcal F)}^2 \nonumber\\
&& -  \frac{1}{32}(\mathcal F^T)_\mu{}^\rho \mathcal M_0\,\mathcal F_{\rho\nu} (\mathcal F^T)^{\mu\sigma} \mathcal M_0\,\mathcal F_{\sigma}{}^{ \nu} {Tr(\mathcal F^T\mathcal M_0\,\mathcal F)}+\dots\,.
\end{eqnarray}

As was mentioned earlier, adding the $T\bar{T}$ deformation to the free Maxwell Lagrangian in four dimensions matches the $F^4$ terms of the BI action expansion. On the other hand, it has been shown \cite{Garousi:2011vs} that the leading order terms of the scattering amplitude of four gauge fields are reproduced by the $F^4$ terms of the BI action. The on-shell BI action is invariant under the linear S-duality up to $F^4$ terms \cite{Babaei-Aghbolagh:2013hia,Garousi:2017fbe}. The $SL(2,R)$ invariant structure of this amplitude agrees with eq.~\eqref{OT2}. It could be uniquely written eq.~\eqref{OT2} in the following appropriate trace terms
 \begin{equation}
 \label{OT22}
 O_{T^2}^{[1/2]}=\frac{1}{2} {Tr(\mathcal F^T\mathcal M_0\,\mathcal F)} -  \frac{1}{64} {Tr(\mathcal F^T\mathcal M_0\,\mathcal F)}^2 +\dots,
 \end{equation}
 where we have used the following identity that holds for $SL(2,R)$ invariant structure of BI theory
 \begin{eqnarray}
 (\mathcal F^T)_\mu{}^\rho \mathcal M_0\,\mathcal F_{\rho\nu} (\mathcal F^T)_\mu{}^\sigma \mathcal M_0\,\mathcal F_{\sigma \nu} =2 {Tr(\mathcal F^T\mathcal M_0\,\mathcal F)} +  \frac{1}{2} {Tr(\mathcal F^T\mathcal M_0\,\mathcal F)}^2.
 \end{eqnarray}
\subsection{S-duality of electrodynamic theories in $D\!=\!2n$}
\label{sec43}
In addition, we are interested in gauge theories of abelian $p$-form potentials which have duality invariant structures. Though the Lagrangian of $p$-form potentials, i.e. $L(P)$, is not invariant under duality transformation, one can introduce a new Lagrangian in $D\!=\!2n$ dimensions (for even $n$) as being invariant under this transformation. We can find this kind of Lagrangian by using a Legendre transformation \cite{Gaillard:1997rt,Aschieri:2008ns,Buratti:2019cbm} that is not a symmetry transformation.

Consider a nonlinear theory of $p$-form potential $F$ with Lagrangian $L(P)$ where $P$ is the Lorentz invariant parameter. The equations of motion and the Bianchi identity for $F$ can be derived from the Lagrangian
\begin{equation}
\label{LD}
L_{D} =  L(P) +\frac{1}{D} F_M G^M,
\end{equation}
where $F_M\equiv F_{\mu_1\dots \mu_n}$ and $G_{M}$ is the field strength of a Lagrange multiplier S-dual potential $A_{\mu_2\dots \mu_n}$. Therefore, the $SL(2,R)$ invariant action is given by
\begin{equation}
\label{actd}
S_{\rm D}=\int L_D \,d^D x=\int\left( L(P) +\frac{1}{D} F_M G^M\right)d^D x.
\end{equation}

Now from the relations \eqref{L1}, \eqref{inv1} and \eqref{inv2} we can determine the flow equation of the action as
\begin{equation}
\label{TTD}
L_{D}=\frac{1}{D}  {T_\mu}^\mu = -  \lambda \frac{\partial L(P)}{\partial \lambda},
\end{equation}
where in 4D we have $L_D=L_{inv}$ which means the invariance of the Lagrangian under $SL(2,R)$ symmetry.

According to the relations \eqref{sl2r2} and \eqref{OT22}, we have obtained an $SL(2,R)$ form for $T\bar{T}$ operators computed in section \ref{sec3} as following. We are interested in a similar structure in arbitrary $D\!=\!2n$  dimensions which is invariant under $\tilde{F}\rightarrow G$ and $G\rightarrow -\tilde{F}$ transformations. In this respect, we introduce an $SL(2,R)$ invariant tensor as follows
\begin{equation} \label{invt}
W_{\mu \nu \alpha \beta}= ( \tilde{F}_{\mu \nu} \tilde{F}_{\alpha \beta })+( G_{\mu \nu} G_{\alpha \beta}).
\end{equation}
It is worth mentioning that though we set $\lambda=1$ for 4D case in subsection \ref{sec41}, the trace of $W$ starts from $\lambda$ while $W^2$ is at the order of $\lambda^2$. The $T\bar{T}$ operator in $D\!=\!2n$ dimensions could be written as
\begin{equation} \label{TTinv}
O_{T^2}=-\frac{1}{n!} {W_{\mu \nu}}^{ \mu \nu}+O(W^2)+\dots,
\end{equation}
where the deformed Lagrangian for each dimension is given by
\begin{equation} \label{geninv}
L_{\lambda}=L_{free}+\frac{1}{8} \int \big(- \frac{1}{n!} {W_{\mu \nu}}^{ \mu \nu}+O(W^2)+\ldots \big) d\lambda.
\end{equation}
The contribution of the term $W$ in the deformed action starts from order $\lambda^2$ or $\mathcal{O}(F^6)$, while the term $W^2$ gives the contribution of order $\lambda^3$ or $\mathcal{O}(F^8)$ to the action. It should be noted that the relation \eqref{geninv} gives the relations in \eqref{gen1} and \eqref{gen2} in $D\!=\!2n$ with exactly specified coefficients in each dimension.
\subsection{$\mathcal{N}=2$ supersymmetric model}
\label{sec44}
The $\mathcal N\!=\!2$ supersymmetric extensions of the BI theory and their duality properties have been found in refs.~\cite{Ketov:1998ku,Ketov:1998sx,Kuzenko:2000uh,Kuzenko:2000tg}. Since the equations of motion in these theories receive contributions from the deformation terms, the NGZ identity appears in some modified form. This implies, at the quantum level, that duality transformations receive modifications. At the quantum level, it has been shown \cite{Carrasco:2011jv} that by considering higher order deformations, maintaining the action’s duality covariance, a theory can preserve the classical duality transformations, at the presence of a duality-invariant counterterm.

Here, we would like to find the $T\bar{T}$ operator of $\mathcal N=2$ supersymmetric theory with $U(1)$-type duality in $D\!=\!4$. It was shown in ref.~\cite{Ferko:2019oyv} that the certain $\mathcal N=2$ deformed theory in $D\!=\!2$ possesses additional non-linearly realized supersymmetries. From this, the $T\bar{T}$ operator of $\mathcal N=1$ BI theory was found in four dimensions.

The action of $\mathcal N=2$ supersymmetric extension of BI theory includes explicit spacetime superfield derivatives as well as spinorial derivatives of the superfields. It is described in an effective framework that is parameterized by four bosonic (Lorentz vectors) and eight fermionic (Lorentz Weyl spinors) coordinates ${\mathcal Z}^A$. It has been proposed in ref.~\cite{Kuzenko:2000uh} that a BI action which exhibits $D_3$-brane type shift symmetry is described by
\begin{eqnarray}
\label{Sact}
S_{BI}&=&\frac18\left(\int d^8\mathcal{Z} \mathcal {W}^2+\int d^8\bar{\mathcal Z} \,\bar{\mathcal W}^2\right)\nonumber\\
&+&\frac{1}{8}\int d^{12}\mathcal {Z}\bigg\{ \mathcal{W}^2\bar{\mathcal W}^2\bigg[\lambda+\frac{\lambda^2}{2}\left({\mathcal D}^4{\mathcal W}^4+\bar{\mathcal D}^4\,\bar{\mathcal W}^4\right)\nonumber\\
&+&\frac{\lambda^3}{4}\Big(({\mathcal D}^4 {\mathcal W}^2)^2+(\bar{\mathcal D}^4\,\bar{\mathcal W}^2)^2+3\,({\mathcal D}^4\mathcal W^2)(\bar{\mathcal D}^4\,\bar{\mathcal W}^2)\Big)\bigg]\nonumber\\
&+&\frac13 \left[\frac{\lambda^2}{3}\mathcal W^3\Box\bar{\mathcal W}^3+\frac{\lambda^3}{2}\Big((\mathcal W^3\Box\bar{\mathcal W}^3) \bar{\mathcal D}^4\,\bar{\mathcal W}^2+(\bar{\mathcal W}^3\Box\mathcal W^3)\mathcal D^4\mathcal W^2+\frac{1}{24}\mathcal W^4\Box^2\bar{\mathcal W}^4\Big)\right]\nonumber\\
&+&{\mathcal O}({\mathcal W}^{10})\bigg\},
\end{eqnarray}
where $\mathcal W(\bar{\mathcal W})$ and $\mathcal D (\bar{\mathcal D})$ are  chiral (anti-chiral) superfield strength and super derivative, respectively. This action is a solution of supersymmetric NGZ condition that is solved perturbatively in the number of fields. The first line in \eqref{Sact} is $\mathcal N=2$ supersymmetric Maxwell action ($S_{free}$) and the other integral is the interacting action ($S_{int}$). The term at the order of $\mathcal O(\lambda)$ produces the known $F^4$ BI action. From the duality transformation for $\mathcal N=2$ supersymmetric theories that proposed in the path integral as a Legendre transform \cite{Broedel:2012gf}, the invariant action of such theories is given by
\begin{eqnarray}\label{Sact2}
S_{inv}=S(\mathcal W,\bar{\mathcal W})-\frac{i}{8}\int d^8{\mathcal Z}\mathcal W\mathcal M+\frac{i}{8}\int d^8\bar{\mathcal Z} \,\bar{\mathcal W}\,\bar{\mathcal M},
\end{eqnarray}
where $\mathcal M$ and $\bar{\mathcal M}$ (similar to the field strength $G$ in non-supersymmetric case) can be defined as $\mathcal M=-4i\frac{\delta}{\delta\mathcal W}S(\mathcal W,\bar{\mathcal W})$ and $\bar{\mathcal M}=4i\frac{\delta}{\delta\bar{\mathcal W}}S(\mathcal W,\bar{\mathcal W})$. These superfield strengths satisfy the Bianchi identities $\mathcal D\mathcal W=\bar{\mathcal D}\,\bar{\mathcal W}$ and $\mathcal D\mathcal M=\bar{\mathcal D}\,\bar{\mathcal M}$.

Using these considerations, we find the invariant action \eqref{Sact2} of a supersymmetric BI theory in terms of superfield strengths and their derivatives as a power series in the coupling constant $\lambda$. Then, the corresponding $T\bar{T}$ operator could be found from eq.~\eqref{inv2} like
\begin{eqnarray}\label{Osusy}
\mathcal O_{T^2}  &=& \int d^{12} \mathcal  Z\,\Bigg\{ {\mathcal W}^2\,\bar{ \mathcal W}^2\,  \Bigg[ 1 +\lambda \, \Big( {\mathcal D}^4 {\mathcal W}^2 + \bar{ \mathcal D}^4 \bar {\mathcal W}^2 \Big)\nonumber\\
&+& \frac{3 \lambda^2}{4} \,\Big( ({\mathcal D}^4 {\mathcal W}^2)^2 + (\bar {\mathcal D}^{4} \bar {\mathcal W}^2)^2
+ 3\, ({\mathcal D}^4 {\mathcal W}^2)(\bar{\mathcal D}^{4} \bar{ \mathcal W}^2)\Big)\Bigg]\nonumber\\
&+&{1 \over3} \,\Bigg[\frac{2\lambda}{3} {\mathcal W}^3 \Box \bar {\mathcal W}^3+\frac{3 \lambda^2}{2}  ({\mathcal W}^3 \Box \bar {\mathcal W}^3) \bar {\mathcal D}^{ 4} \bar {\mathcal W}^2+ (\bar {\mathcal W}^3 \Box {\mathcal W}^3) {\mathcal D}^4 {\mathcal W}^2+{ 1 \over 24} {\mathcal W}^4 \Box^2 \bar {\mathcal W}^4  \Bigg]\nonumber\\
&+&{\mathcal O}({\mathcal W}^{10})  \Bigg\}.
\end{eqnarray}
The $\mathcal N=2$ supersymmetric action could be presented in the general form\cite{Broedel:2012gf}
\begin{equation}
\label{Sact3}
S_{\mathcal N=2}=S_{free}+\int d^{12}\mathcal Z\mathcal W^2\bar{\mathcal W}^2\,\mathcal Y(\mathcal D^4\mathcal W^2,\bar{\mathcal D}^4 \bar{\mathcal W}^2)+\mathcal O(\partial_{\mu}\mathcal W),
\end{equation}
where $\mathcal Y$ is a BI-type functional. Finding the invariant action up to $\mathcal O(\partial_{\mu}\mathcal W)$ from $S_{inv}=-\lambda\frac{\partial S}{\partial\lambda}$ , we can determine the $T\bar{T}$ operator in general form using eq.~\eqref{inv2}. Thus, we obtain
\begin{equation}
\label{TTsusy}
O_{T^2} =
8\,\mathcal W^2\,\bar{ \mathcal W}^2 \frac{d}{d\lambda} {\mathcal Y}\; (\mathcal D^4 {\mathcal  W}^2,\ \bar{\mathcal D}^4 \bar {\mathcal W}^2) .
\end{equation}

It was shown in ref.~\cite{Chemissany:2012pf} that the higher derivative terms $\mathcal O(\partial_{\mu}\mathcal W)$ in eq.~\eqref{Sact3} reduce to the BI action at order $\mathcal O(\partial F)^4$ in the non-supersymmetric level that come from the one-loop amplitude of four gauge fields. These higher derivative terms respect to NGZ identity and have found in the form that is manifestly $SL(2,R)$ invariant \cite{BabaeiVelni:2019ptj}. This implies that one can find the invariant higher derivative action which corresponds to $\mathcal O(\partial_{\mu}\mathcal W)$ and obtain the corresponding $T\bar{T}$ operator that contributes to eq.~\eqref{TTsusy}.
\section{Conclusion and outlook}
\label{sec5}

The interest in studying a class of QFTs perturbed by irrelevant $T\bar{T}$ operators is one of the most important subjects in recent research arena. Among the infinite number of possible perturbations of a given QFT,
the latter operator displays very special and universal features. These perturbations may lead to singular RG flows where the UV fixed point is not well-defined. Although many studies have been done in the case of 2D QFTs, however a little attention have been dedicated to investigate this deformation in higher-dimensional QFTs. It is worth to emphasize that there are no conformal theories at the quantum level for $D>6$, but the generalization of $T\bar{T}$ operator for non-conformal QFTs may open new windows on gauge/gravity duality regardless of conformal symmetry. Due to this fact, in this paper we studied the construction of $T\bar{T}$ deformation for some non-linear electrodynamic BI-type theories in general $D(=2n)$-dimensional spacetime.

We have proposed a $T\bar{T}$ operator $O_{T^2}$ given by \eqref{TT3} in $2n$ dimensions for even $n$. This kind of deformation was compatible with the results reported in 2D and 4D QFTs, but was in conflict with the Taylor's proposal \cite{Taylor:2018xcy} in general dimensions. With confidence to this proposal we reproduced the expansion of BI-type theories in \eqref{exp1} by deforming the free Lagrangian by this $T\bar{T}$ operator in 2 and 4 dimensions. We observed that not only the EM tensor \eqref{EMT1} satisfies the flow equation \eqref{TEM} for each $2n$-dimensional theory, but also helps us to reconstruct the deformed action through $O^{[1/n]}_{T^2}$ in \eqref{TT3}.

But in higher dimensions $D\geq8$, the compatibility between the BI expansion and the deformation of free theory is threatened by the existence of a term like $K_2$ which can not be written only in terms of Lorentz invariant variables $P_1$ and $P_2$ in each dimension. Therefore, we trusted to our proposal to deform the free action of higher dimensional electrodynamic theory at any arbitrary order of the deformation parameter.
In this deformation we showed that the contribution of the second term in $T\bar{T}$ operator, i.e. $(T_{\mu}{}^{\mu})^2$, starts from $\lambda^3$ in the deformed Lagrangian. We have also shown that one can find a general form for the deformed Lagrangian at each order of $\lambda$ which at the order of $\lambda$ and $\lambda^2$ are given respectively by the relations \eqref{gen1} and \eqref{gen2}.

Following the general strategy suggested in the text, one can also deform the free theory of $p$-form gauge theories in diverse $D=2n+2$ dimensions. For instance, we deform a 6D free theory of 3-form self-interacting field strength, i.e. $\frac{1}{12} F_{ \mu\nu\rho} F^{\mu\nu\rho}$, but the results is incompatible with the expansion of the  6D duality-invariant non-linear BI theory \cite{Bandos:2020jsw}. It has been shown in ref.~\cite{Bandos:2020hgy} that in 6D there is a unique non-linear conformal modification of the free chiral 2-form theory which is related to a non-linear modification of 4D Maxwell electrodynamics by dimensional reduction (see also \cite{Berman:1997iz}). It would be of interest to investigate the $T\bar{T}$ deformation for a 6D non-linear theory of the so-called chiral 2-forms whose 3-form field strength satisfies a self-duality condition.

According to the Cardy's proposal in ref.~\cite{Cardy:2018sdv}, one can define the $T\bar{T}$ operator as $O_{T^2}\sim a T^{ \mu \nu}T_{ \mu \nu}+ b {T_\mu}^\mu {T_\nu}^\nu$ where the constant coefficients $a$ and $b$ in this proposal depend on the corresponding theory. For example, we observed that the coefficient $b$ is equal to $1/(D-1)$ for gravitational theory in general dimension \cite{Taylor:2018xcy} while from our calculations in this paper, one finds that it is $1/n$ for $D(=2n)$-dimensional gauge theories. Also, it has been shown in refs.~\cite{Schwarz:1982jn,Gross:1986iv} that the effective action of type II superstring theory, which was obtained from deformation of type II supergravity theory, starts from order $\lambda^3 \sim\alpha'^3$ as
\begin{eqnarray}\label{ss1}
S_{eff}&=& S_0+ \alpha^{\prime 3} S_3+\dots\, .
\end{eqnarray}
In fact, in the $Ramond\!-\!Ramond$ sector of type IIB superstring theory, there is a term similar to Maxwell theory in $D\!=\!10$ dimensions which is constructed from $5$-forms as follows
\begin{eqnarray}\label{ss2}
S_{0}&=&- \tfrac{1}{2\times 5!}\int d^{10}x \,\,\, F^{\mu \alpha\gamma \iota \kappa } F_{\mu \alpha\gamma \iota \kappa }.
\end{eqnarray}
From the relation \eqref{ss1}, the deformation of action \eqref{ss2} in the $Ramond-Ramond$ sector starts from $\lambda^3 \sim\alpha'^3$. Thus, as shown in section \ref{sec3}, since the contribution of term $T^{ \mu \nu}T_{ \mu \nu}$ starts from $\lambda$ and $(T_{\mu}{}^{\mu})^2$ from $\lambda^3$, it would be logical to expect that the deformation of type IIB superstring theory with $T\bar{T}$ operator in the $Ramond\!-\!Ramond$ sector is consistent with the assumption $a=0$ in $O_{T^2}$.

We have also investigated the structure of $T\bar{T}$ operator for a class of non-linear electrodynamic BI-type theories in 4D. In particular, we considered a gauge theory with a dilatonic field as gauge coupling and showed that the deformation of free theory made by \eqref{TT3} is also consistent with the expansion \eqref{exp2}. It was shown that this operator inherits the $SL(2,R)$ invariant symmetry of the theory though the action is not.  We generalized this implication to higher dimensional electrodynamic theories investigated in section \ref{sec3} by introducing an $SL(2,R)$ invariant tensor and recast the $T\bar{T}$ operator and deformed Lagrangian in terms of this tensor in eqs.~\eqref{TTinv} and \eqref{geninv}.
As a toy model in the context of supersymmetry, we constructed the $T\bar{T}$ operator for $\mathcal N=2$ supersymmetric BI theory in eq.~\eqref{TTsusy} from proposals \eqref{inv1} and \eqref{inv2} for the functional $\mathcal Y$ in the space of superfields and superderivatives.

It would be of interest to study the $T\bar{T}$ operator in the case of $\mathcal N=4$ supersymmetric theory in 4D. The $\mathcal N=4$ action has a maximal number of supersymmetries that has been proposed in ref.~\cite{Bergshoeff:2013pia}. It could be found the corresponding $SL(2,R)$ invariant action and then, the $T\bar{T}$ operator using eq.~\eqref{inv2}. We leave the details of this interesting issue for future study.

\section*{Acknowledgment}
We would like to thank S. Sethi, M. Alishahiha, M. R. Garousi and G. Jafari for valuable comments and discussions. We are also grateful to D. Sorokin for fruitful discussion and helpful comments on the generalized BI theories in $D=6$ dimensions.


\begin{thebibliography}{99}

\bibitem{Zamolodchikov:2004ce}
A.~B.~Zamolodchikov,
``Expectation value of composite field T anti-T in two-dimensional quantum field theory,''
[arXiv:hep-th/0401146 [hep-th]].

\bibitem{Smirnov:2016lqw}
F.~A.~Smirnov and A.~B.~Zamolodchikov,
``On space of integrable quantum field theories,''
Nucl. Phys. B \textbf{915}, 363-383 (2017),
[arXiv:1608.05499 [hep-th]].

\bibitem{Cavaglia:2016oda}
A.~Cavagli\`a, S.~Negro, I.~M.~Sz\'ecs\'enyi and R.~Tateo,
``$T \bar{T}$-deformed 2D Quantum Field Theories,''
JHEP \textbf{10}, 112 (2016),
[arXiv:1608.05534 [hep-th]].

\bibitem{Guica:2017lia}
M.~Guica,
``An integrable Lorentz-breaking deformation of two-dimensional CFTs,''
SciPost Phys. \textbf{5}, no.5, 048 (2018),
[arXiv:1710.08415 [hep-th]].

\bibitem{Chen:2018keo}
C.~Chen, P.~Conkey, S.~Dubovsky and G.~Hern\'andez-Chifflet,
``Undressing Confining Flux Tubes with $T\bar T$,''
Phys. Rev. D \textbf{98}, no.11, 114024 (2018),
[arXiv:1808.01339 [hep-th]].

\bibitem{Aharony:2018bad}
O.~Aharony, S.~Datta, A.~Giveon, Y.~Jiang and D.~Kutasov,
``Modular invariance and uniqueness of $T\bar{T}$ deformed CFT,''
JHEP \textbf{01}, 086 (2019),
[arXiv:1808.02492 [hep-th]].

\bibitem{Cardy:2018sdv}
J.~Cardy,
``The $ T\overline{T} $ deformation of quantum field theory as random geometry,''
JHEP \textbf{10}, 186 (2018),
[arXiv:1801.06895 [hep-th]].

\bibitem{Datta:2018thy}
S.~Datta and Y.~Jiang,
``$T\bar{T}$ deformed partition functions,''
JHEP \textbf{08}, 106 (2018),
[arXiv:1806.07426 [hep-th]].

\bibitem{Bonelli:2018kik}
G.~Bonelli, N.~Doroud and M.~Zhu,
``$T \bar{T}$-deformations in closed form,''
JHEP \textbf{06}, 149 (2018),
[arXiv:1804.10967 [hep-th]].

\bibitem{Brennan:2019azg}
T.~D.~Brennan, C.~Ferko and S.~Sethi,
``A Non-Abelian Analogue of DBI from $T \overline{T}$,''
SciPost Phys. \textbf{8}, no.4, 052 (2020),
[arXiv:1912.12389 [hep-th]].

\bibitem{McGough:2016lol}
L.~McGough, M.~Mezei and H.~Verlinde,
``Moving the CFT into the bulk with $ T\overline{T} $,''
JHEP \textbf{04}, 010 (2018),
[arXiv:1611.03470 [hep-th]].

\bibitem{Kraus:2018xrn}
P.~Kraus, J.~Liu and D.~Marolf,
``Cutoff AdS$_{3}$ versus the $ T\overline{T} $ deformation,''
JHEP \textbf{07}, 027 (2018),
[arXiv:1801.02714 [hep-th]].

\bibitem{Heemskerk:2009pn}
I.~Heemskerk, J.~Penedones, J.~Polchinski and J.~Sully,
``Holography from Conformal Field Theory,''
JHEP \textbf{10}, 079 (2009),
[arXiv:0907.0151 [hep-th]].

\bibitem{Giribet:2017imm}
G.~Giribet,
``$T\bar{T}$-deformations, AdS/CFT and correlation functions,''
JHEP \textbf{02}, 114 (2018),
[arXiv:1711.02716 [hep-th]].

\bibitem{Giveon:2017myj}
A.~Giveon, N.~Itzhaki and D.~Kutasov,
``A solvable irrelevant deformation of AdS$_{3}$/CFT$_{2}$,''
JHEP \textbf{12}, 155 (2017),
[arXiv:1707.05800 [hep-th]].

\bibitem{Aharony:2018vux}
O.~Aharony and T.~Vaknin,
``The TT* deformation at large central charge,''
JHEP \textbf{05}, 166 (2018),
[arXiv:1803.00100 [hep-th]].

\bibitem{Donnelly:2018bef}
W.~Donnelly and V.~Shyam,
``Entanglement entropy and $T \overline{T}$ deformation,''
Phys. Rev. Lett. \textbf{121}, no.13, 131602 (2018),
[arXiv:1806.07444 [hep-th]].

\bibitem{Hartman:2018tkw}
T.~Hartman, J.~Kruthoff, E.~Shaghoulian and A.~Tajdini,
``Holography at finite cutoff with a $T^2$ deformation,''
JHEP \textbf{03}, 004 (2019),
[arXiv:1807.11401 [hep-th]].

\bibitem{Conti:2018tca}
R.~Conti, S.~Negro and R.~Tateo,
``The $ \mathrm{T}\overline{\mathrm{T}} $ perturbation and its geometric interpretation,''
JHEP \textbf{02}, 085 (2019),
[arXiv:1809.09593 [hep-th]].

\bibitem{Taylor:2018xcy}
M.~Taylor,
``TT deformations in general dimensions,''
[arXiv:1805.10287 [hep-th]].

\bibitem{Maldacena:1997re}
J.~M.~Maldacena,
``The Large N limit of superconformal field theories and supergravity,''
Int. J. Theor. Phys. \textbf{38}, 1113-1133 (1999),
[arXiv:hep-th/9711200 [hep-th]].

\bibitem{Conti:2018jho}
R.~Conti, L.~Iannella, S.~Negro and R.~Tateo,
``Generalised Born-Infeld models, Lax operators and the $ \mathrm{T}\overline{\mathrm{T}} $ perturbation,''
JHEP \textbf{11}, 007 (2018),
[arXiv:1806.11515 [hep-th]].

\bibitem{Ferko:2019oyv}
C.~Ferko, H.~Jiang, S.~Sethi and G.~Tartaglino-Mazzucchelli,
``Non-linear supersymmetry and $ T\overline{T} $-like flows,''
JHEP \textbf{02}, 016 (2020),
[arXiv:1910.01599 [hep-th]].

\bibitem{Betzios:2020sro}
P.~Betzios, E.~Kiritsis and V.~Niarchos,
``Emergent gravity from hidden sectors and TT deformations,''
[arXiv:2010.04729 [hep-th]].

\bibitem{Nastase:2021mgy}
H.~Nastase and J.~Sonnenschein,
``A $T\bar T$-like deformation of the Skyrme model and the Heisenberg model of nucleon-nucleon scattering,''
[arXiv:2101.08232 [hep-th]].

\bibitem{Frolov:2021oav}
S.~Frolov and C.~Esper,
``$T\overline{T}$ Deformations of nonrelativistic models,''
[arXiv:2102.12435 [hep-th]].

\bibitem{Gibbons:1995ap}
G.~W.~Gibbons and D.~A.~Rasheed,
``Sl(2,R) invariance of nonlinear electrodynamics coupled to an axion and a dilaton,''
Phys. Lett. B \textbf{365}, 46-50 (1996),
[arXiv:hep-th/9509141 [hep-th]].

\bibitem{Gaillard:1997rt}
M.~K.~Gaillard and B.~Zumino,
``Nonlinear electromagnetic selfduality and Legendre transformations,''
[arXiv:hep-th/9712103 [hep-th]].

\bibitem{BabaeiVelni:2016qea}
K.~Babaei Velni and H.~Babaei-Aghbolagh,
``On SL (2,R) symmetry in nonlinear electrodynamics theories,''
Nucl. Phys. B \textbf{913}, 987-1000 (2016),
[arXiv:1610.07790 [hep-th]].

\bibitem{Aschieri:2008ns}
P.~Aschieri, S.~Ferrara and B.~Zumino,
``Duality Rotations in Nonlinear Electrodynamics and in Extended Supergravity,''
Riv. Nuovo Cim. \textbf{31}, 625-708 (2008),
[arXiv:0807.4039 [hep-th]].

\bibitem{Aschieri:2013nda}
P.~Aschieri and S.~Ferrara,
``Constitutive relations and Schroedinger's formulation of nonlinear electromagnetic theories,''
JHEP \textbf{05}, 087 (2013),
[arXiv:1302.4737 [hep-th]].

\bibitem{Gaillard:1981rj}
M.~K.~Gaillard and B.~Zumino,
``Duality Rotations for Interacting Fields,''
Nucl. Phys. B \textbf{193}, 221-244 (1981).

\bibitem{Green:1996qg}
M.~B.~Green and M.~Gutperle,
``Comments on three-branes,''
Phys. Lett. B \textbf{377}, 28-35 (1996),
[arXiv:hep-th/9602077 [hep-th]].

\bibitem{Baggio:2018rpv}
M.~Baggio, A.~Sfondrini, G.~Tartaglino-Mazzucchelli and H.~Walsh,
``On $ T\overline{T} $ deformations and supersymmetry,''
JHEP \textbf{06}, 063 (2019),
[arXiv:1811.00533 [hep-th]].

\bibitem{Chang:2018dge}
C.~K.~Chang, C.~Ferko and S.~Sethi,
``Supersymmetry and $ T\overline{T} $ deformations,''
JHEP \textbf{04}, 131 (2019),
[arXiv:1811.01895 [hep-th]].

\bibitem{Jiang:2019hux}
H.~Jiang, A.~Sfondrini and G.~Tartaglino-Mazzucchelli,
``$T\bar{T}$ deformations with $\mathcal{N}=(0,2)$ supersymmetry,''
Phys. Rev. D \textbf{100}, no.4, 046017 (2019),
[arXiv:1904.04760 [hep-th]].

\bibitem{Chang:2019kiu}
C.~K.~Chang, C.~Ferko, S.~Sethi, A.~Sfondrini and G.~Tartaglino-Mazzucchelli,
``$T\bar{T}$ flows and (2,2) supersymmetry,''
Phys. Rev. D \textbf{101}, no.2, 026008 (2020),
[arXiv:1906.00467 [hep-th]].

\bibitem{Kuzenko:2000uh}
S.~M.~Kuzenko and S.~Theisen,
``Nonlinear selfduality and supersymmetry,''
Fortsch. Phys. \textbf{49}, 273-309 (2001),
[arXiv:hep-th/0007231 [hep-th]].

\bibitem{Broedel:2012gf}
J.~Broedel, J.~J.~M.~Carrasco, S.~Ferrara, R.~Kallosh and R.~Roiban,
``N=2 Supersymmetry and U(1)-Duality,''
Phys. Rev. D \textbf{85}, 125036 (2012),
[arXiv:1202.0014 [hep-th]].

\bibitem{Brennan:2020dkw}
T.~D.~Brennan, C.~Ferko, E.~Martinec and S.~Sethi,
``Defining the $T \overline{T}$ Deformation on $\mathrm{AdS}_2$,''
[arXiv:2005.00431 [hep-th]].

\bibitem{Rosenhaus:2019utc}
V.~Rosenhaus and M.~Smolkin,
``Integrability and renormalization under $T \bar T$,''
Phys. Rev. D \textbf{102}, no.6, 065009 (2020),
[arXiv:1909.02640 [hep-th]].

\bibitem{Buratti:2019cbm}
G.~Buratti, K.~Lechner and L.~Melotti,
``Duality invariant self-interactions of abelian p-forms in arbitrary dimensions,''
JHEP \textbf{09}, 022 (2019),
[arXiv:1906.07094 [hep-th]].

\bibitem{Buratti:2019guq}
G.~Buratti, K.~Lechner and L.~Melotti,
``Self-interacting chiral p-forms in higher dimensions,''
Phys. Lett. B \textbf{798}, 135018 (2019),
[arXiv:1909.10404 [hep-th]].

\bibitem{Henningson:1998gx}
M.~Henningson and K.~Skenderis,
``The Holographic Weyl anomaly,''
JHEP \textbf{07}, 023 (1998),
[arXiv:hep-th/9806087 [hep-th]].

\bibitem{Brown:1986nw}
J.~D.~Brown and M.~Henneaux,
``Central Charges in the Canonical Realization of Asymptotic Symmetries: An Example from Three-Dimensional Gravity,''
Commun. Math. Phys. \textbf{104}, 207-226 (1986).

\bibitem{Bastianelli:2000hi}
F.~Bastianelli, S.~Frolov and A.~A.~Tseytlin,
``Conformal anomaly of (2,0) tensor multiplet in six-dimensions and AdS / CFT correspondence,''
JHEP \textbf{02}, 013 (2000),
[arXiv:hep-th/0001041 [hep-th]].

\bibitem{Bandos:2020hgy}
I.~Bandos, K.~Lechner, D.~Sorokin and P.~K.~Townsend,
``On p-form gauge theories and their conformal limits,''
[arXiv:2012.09286 [hep-th]].

\bibitem{Bandos:2020jsw}
I.~Bandos, K.~Lechner, D.~Sorokin and P.~K.~Townsend,
``A non-linear duality-invariant conformal extension of Maxwell's equations,''
Phys. Rev. D \textbf{102}, 121703 (2020),
[arXiv:2007.09092 [hep-th]].

\bibitem{Rasheed:1997ns}
D.~A.~Rasheed,
``Nonlinear electrodynamics: Zeroth and first laws of black hole mechanics,''
[arXiv:hep-th/9702087 [hep-th]].

\bibitem{Garousi:2011vs}
M.~R.~Garousi,
``On S-duality of D$_3$-brane S-matrix,''
Phys. Rev. D \textbf{84}, 126019 (2011),
[arXiv:1108.4782 [hep-th]].

\bibitem{Babaei-Aghbolagh:2013hia}
H.~Babaei-Aghbolagh and M.~R.~Garousi,
``S-duality of tree-level S-matrix elements in D3-brane effective action,''
Phys. Rev. D \textbf{88}, no.2, 026008 (2013),
[arXiv:1304.2938 [hep-th]].

\bibitem{Garousi:2017fbe}
M.~R.~Garousi,
``Duality constraints on effective actions,''
Phys. Rept. \textbf{702}, 1-30 (2017),
[arXiv:1702.00191 [hep-th]].

\bibitem{Ketov:1998ku}
S.~V.~Ketov,
``A Manifestly N=2 supersymmetric Born-Infeld action,''
Mod. Phys. Lett. A \textbf{14}, 501-510 (1999),
[arXiv:hep-th/9809121 [hep-th]].

\bibitem{Ketov:1998sx}
S.~V.~Ketov,
``Born-Infeld-Goldstone superfield actions for gauge fixed D-5 branes and D-3 branes in 6-d,''
Nucl. Phys. B \textbf{553}, 250-282 (1999),
[arXiv:hep-th/9812051 [hep-th]].

\bibitem{Kuzenko:2000tg}
S.~M.~Kuzenko and S.~Theisen,
``Supersymmetric duality rotations,''
JHEP \textbf{03}, 034 (2000),
[arXiv:hep-th/0001068 [hep-th]].

\bibitem{Carrasco:2011jv}
J.~J.~M.~Carrasco, R.~Kallosh and R.~Roiban,
``Covariant procedures for perturbative non-linear deformation of duality-invariant theories,''
Phys. Rev. D \textbf{85}, 025007 (2012),
[arXiv:1108.4390 [hep-th]].

\bibitem{Chemissany:2012pf}
W.~Chemissany, S.~Ferrara, R.~Kallosh and C.~S.~Shahbazi,
``N=2 Supergravity Counterterms, Off and On Shell,''
JHEP \textbf{12}, 089 (2012),
[arXiv:1208.4801 [hep-th]].

\bibitem{BabaeiVelni:2019ptj}
K.~Babaei Velni and H.~Babaei-Aghbolagh,
``$S$-dual amplitude and $D_3$-brane couplings,''
Phys. Rev. D \textbf{99}, no.6, 066007 (2019),
[arXiv:1901.00198 [hep-th]].

\bibitem{Berman:1997iz}
D.~Berman,
``SL(2,Z) duality of Born-Infeld theory from nonlinear selfdual electrodynamics in six-dimensions,''
Phys. Lett. B \textbf{409}, 153-159 (1997),
[arXiv:hep-th/9706208 [hep-th]].

\bibitem{Schwarz:1982jn}
J.~H.~Schwarz,
``Superstring Theory,''
Phys. Rept. \textbf{89}, 223-322 (1982).

\bibitem{Gross:1986iv}
D.~J.~Gross and E.~Witten,
``Superstring Modifications of Einstein's Equations,''
Nucl. Phys. B \textbf{277}, 1 (1986).

\bibitem{Bergshoeff:2013pia}
E.~Bergshoeff, F.~Coomans, R.~Kallosh, C.~S.~Shahbazi and A.~Van Proeyen,
``Dirac-Born-Infeld-Volkov-Akulov and Deformation of Supersymmetry,''
JHEP \textbf{08}, 100 (2013),
[arXiv:1303.5662 [hep-th]].

\end{thebibliography}
\end{document}